\documentclass{elsart}

\usepackage{epsf}
\usepackage[dvips]{epsfig}
\usepackage[T1]{fontenc}
\usepackage[latin1]{inputenc}
\usepackage{times,pifont}

\def\simless{\mathbin{\lower 3pt\hbox
   {$\rlap{\raise 5pt\hbox{$\char'074$}}\mathchar"7218$}}}   
\def\simgreat{\mathbin{\lower 3pt\hbox
   {$\rlap{\raise 5pt\hbox{$\char'076$}}\mathchar"7218$}}}
\def\one{{}^{(\!1\!)}}
\def\two{{}^{(\!2\!)}}
\def\meshx#1{#1_{\rm m,x}^{\!\dagger}}
\def\mesh#1{#1_{\rm m}^{\!\dagger}}
\def\dbgx#1{#1_x^{\!\dagger}}
\def\dbg#1{#1^{\!\dagger}}
\def\comb{\raise 0.7ex\hbox{$\scriptscriptstyle{\coprod}$} 
          \kern -0.40em\raise 0.7ex\hbox{$\scriptscriptstyle{\coprod}$}}

\begin{document}

\begin{frontmatter}
  
\title{SUPERBOX -- An Efficient Code for Collisionless Galactic
  Dynamics} 

\author[ARI]{M. Fellhauer \thanksref{e1}}
\author[ITA]{P. Kroupa \thanksref{e2}}
\author[Edin]{H. Baumgardt \thanksref{e3}}
\author[ARI]{R. Bien \thanksref{e1}}
\author[ARI]{C.M. Boily \thanksref{e1}}
\author[ARI]{R. Spurzem \thanksref{e1}}
\author[Bosch]{N. Wassmer}

\address[ARI]{Astronomisches Rechen-Institut,
    M\"onchhofstr.~12-14, D-69120~Heidelberg, Germany}
\address[ITA]{Institut f\"ur Theoretische Astrophysik,
    Universit\"at Heidelberg, Tiergartenstr.~15,
    D-69121~Heidelberg, Germany} 
\address[Edin]{Department of Mathematics and Statistics,
    University of Edinburgh, King's Buildings, Edinburgh EH9 3JZ,
    Scotland (UK)} 
\address[Bosch]{Robert Bosch GmbH, Robert Bosch Str.~1, D-77815
    Buehl, Germany}

\thanks[e1]{e-mail: {mike,reinhold,cmb,spurzem}@ari.uni-heidelberg.de}
\thanks[e2]{e-mail: pavel@ita.uni-heidelberg.de}
\thanks[e3]{e-mail: holger@maths.ed.ac.uk}

\begin{abstract}
  We present {\sc Superbox}, a particle-mesh code with high
  resolution sub-grids and an NGP (nearest grid point)
  force-calculation scheme based on the second derivatives of the
  potential.  {\sc Superbox} implements a fast low-storage
  FFT-algorithm, giving the possibility to work with millions of
  particles on desk-top computers.  Test calculations show energy
  and angular momentum conservation to one part in $10^{5}$ per
  crossing-time.  The effects of grid and numerical relaxation
  remain negligible, even when these calculations cover a
  Hubble-time of evolution.  As the sub-grids follow the
  trajectories of  individual galaxies, the code allows a highly
  resolved treatment of interactions in clusters of galaxies,
  such as high-velocity encounters between elliptical galaxies
  and the tidal disruption of dwarf galaxies.  Excellent
  agreement is obtained in a comparison with a direct-summation
  N-body code running on special-purpose {\sc Grape}~3 hardware.
  The orbital decay of satellite galaxies due to dynamical
  friction obtained with {\sc Superbox} agrees with
  Chandrasekhar's treatment when the Coulomb logarithm $\ln
  \Lambda \approx 1.5$. 
\end{abstract}


\end{frontmatter}


\section{Introduction}
\label{sec:intro}

\noindent
In direct-summation N-body methods, the computational time
$t_{CPU}$ scales with the square of the particle number $N_{\rm
  p}$, $t_{CPU} \sim N_{\rm p}^2$.  This relation inevitably
leads to bottlenecks when computing large-N galaxy models.  While
it is still impossible to model galaxies star by star over
sensible time-laps, alternative algorithms have been developed
over a number of years which ease the computer requirements at
the expense of the exact $N_{\rm p}^2$ summation, yet provide an
adequate treatment of Newtonian galactic dynamics.  Examples of
such techniques are the tree method (Barnes \& Hut 1986, Dav\'e,
Dubinski \& Hernquist 1997), and the multi-grid or nested grid
particle mesh-codes like {\sc Pandora} (Villumsen 1989), {\sc
  Hydra} (Pearce \& Couchman 1997), {\sc Superbox}, or
combinations of the methods like the adaptive refinement tree
({\sc ART}) (Kravtsov, Klypin \& Khokhlov.\ 1997), which can cope
with very inhomogeneous matter distributions and have high
spatial resolution at the density maxima. 

While the hierarchical tree-method is independent of the geometry 
of the problem, its main disadvantage is the limitation in
particle number and the softening needed to avoid undesired
two-body relaxation.  In contrast, the main advantage of a
particle-mesh technique is the ability to use very high particle
numbers (up to several millions on desk-top computers), because
the CPU-time scales only linearly with the number of particles,
and with $N_{\rm gc} \log N_{\rm gc}$, where $N_{\rm gc}$ is the
number of grid-cells.  High particle numbers keep the statistical
noise low.  The disadvantage of a grid-based method, however, is
the dependency on the geometry of the grid.  But with increasing
spatial resolution, the geometry of the cells becomes less
important.  The only limitation in resolving phenomena is the
size of one grid-cell.  A general discussion and comparison of
various numerical schemes will be found in Sellwood (1987), and
-- in the context of direct $N$-body simulations -- in Spurzem
(1999). 

The basic idea behind {\sc Superbox}, as originally conceived by
R. Bien (Bien et al.\ 1991), is to increase the resolution only
at places where it is necessary, while simultaneously keeping the 
computational overhead as small as possible by using a fixed
(i.e.\ not adaptive) nested grid-architecture.  Accuracy is
improved by the use of nested high-resolution sub-grids and a
linear force interpolation to the exact position of the particle
inside a cell.  Two higher-resolution sub-grids are introduced:
the medium-resolution grid contains an entire galaxy, and the
high-resolution grid treats its core.  Both grids stay focused on
the galaxy, moving through the 'local universe' that is contained
in the coarse outermost grid. 

{\sc Superbox} has already been used successfully in a variety of 
investigations, namely in the study of the high velocity
encounter of the two similar early-type galaxies NGC\,4782/4783
by Madejski \& Bien (1993), in the survey of encounters between
elliptical galaxies by Wassmer et al.\ (1993), and in the
research on tidal disruption of satellite dwarf galaxies by
Kroupa (1997) and Klessen \& Kroupa (1998).  A version of {\sc
  Superbox} exists that includes sticky particles as a simple
model of the dynamics of cold molecular gas clouds (Fellhauer
1996). 

In this article we provide an implementation and benchmarks for
the {\sc Superbox} algorithm.  In Section~\ref{sec:method},  we
lay out the mathematics behind the algorithm and give details of
how the potential is mapped on the different grids.  We then
move on to test the code: Section~\ref{sec:conserv} deals with
the conservation of energy and angular momentum, while
Section~\ref{sec:relax} discusses relaxation.
Section~\ref{sec:memcpu} contains profiling information about
storage and CPU-time requirements.  Comparison with a
direct-summation  $N_{\rm p}$-body code is made in
Section~\ref{sec:fin_valid}.  In Section~\ref{sec:dynfric} we
show -- as an example -- the good agreement of satellite decay
with Chandrasekhar's formula for dynamical friction, and close
with conclusions in Section~\ref{sec:conc}.  


\section{Method}
\label{sec:method}

\noindent
Detailed discussions of the particle-mesh (PM) technique can be
found in Eastwood \& Brownrigg (1978), Hockney \& Eastwood (1981)
and Sellwood (1987). 

Briefly, the array of mass densities, $\varrho_{ijk}$, is derived
by counting the number of particles in each Cartesian grid cell
$i,j,k$, i.e. by using the simple particle-in-cell (NGP = ``Nearest
Grid Point'')
algorithm.  All particles belonging to one galaxy $N_{\rm p,gal}$
have the same mass, $m = M_{\rm gal} / N_{\rm p,gal}$.  An
alternative would be to use a cloud-in-cell technique, which is
known to improve momentum and energy conservation, in which the
mass of the particle is distributed proportionally among the
neighbouring cells.  This is, however, not necessary, since the
number of particles is large. 
 
\begin{figure}[ht!]
  \begin{center}
    \leavevmode
    \epsfxsize=10cm
    \epsfysize=13cm
    \epsffile{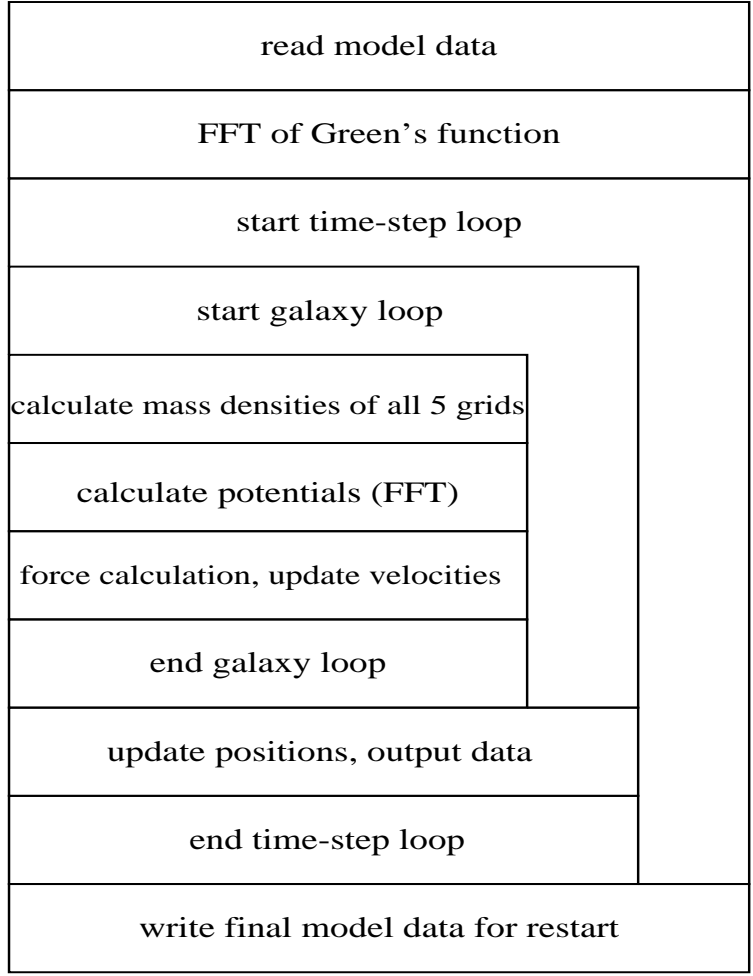}
    \caption{Float-chart for {\sc Superbox}.}
    \label{super}
  \end{center}
\end{figure}

Poisson's equation is solved for this density array to get the
grid-based potential, $\Phi_{ijk}$, which becomes, 
\begin{eqnarray}
  \Phi_{ijk} & = & G \; \sum_{a,b,c=0}^{N-1} \ \varrho_{abc}
  \cdot H_{a-i,b-j,c-k}, \ \  i,j,k=0,...,N-1, \label{f2.0a}
\end{eqnarray}
where $N$ denotes the number of grid-cells per dimension ($ N^{3}
= N_{\rm gc} $), and $H_{ijk}$ is some Green's function.  To
avoid this $N_{\rm gc}^{2}$ procedure, the discrete Fast Fourier
Transform (FFT) is used, for which $N=2^{K}$, $K>0$ being an
integer.  The stationary Green's function is Fourier transformed
once at the beginning of the calculation, and only the density
array is transformed at each time-step:
\begin{eqnarray}
  \hat{\varrho}_{abc} & = & \sum_{i,j,k=0}^{N-1} \; \varrho_{ijk} 
  \cdot \exp \left(- \sqrt{-1}\ \frac{2\pi}{N} \left(ai+bj+ck
  \right) \right), \label{f2.0b}\\ 
  \hat{H}_{abc} & = & \sum_{i,j,k=0}^{N-1} \; H_{ijk} \cdot \exp
  \left( - \sqrt{-1}\ \frac{2\pi}{N} \left( ai+bj+ck \right)
  \right). \nonumber 
\end{eqnarray}
The two resulting arrays are multiplied cell by cell and
transformed back to get the grid-based potential,  
\begin{eqnarray}
   \Phi_{ijk} & = & \frac{G}{N^{3}} \; \sum_{a,b,c=0}^{N-1}
   \hat{\varrho}_{abc} \cdot \hat{H}_{abc} \cdot \exp \left(
   \sqrt{-1}\ \frac{2\pi}{N} \left( ai+bj+ck \right)
   \right). \label{f2.0c} 
\end{eqnarray}
The potential is differentiated numerically to estimate the
accelerations for each particle, each of which is pushed forward
on its orbit using the leapfrog scheme (see Fig.~\ref{super}
and sections below).

\subsection{Green's function and the FFT}
\label{sec:green}
\noindent
The simplest Green's function is adopted, in which the side of a
grid cell has unit length:
\begin{eqnarray}
  H_{ijk} & = & \frac{1}{\sqrt{i^{2}+j^{2}+k^{2}}},\ \ \
  i,j,k=0,1,..,N \label{eq:green_fn} \\ 
  H_{000} & = & 4/3. \nonumber
\end{eqnarray}
The value of $H_{000}$ is somewhat arbitrary, and $H_{000} =
H_{100}$ is often used (see e.g. Sellwood 1987).  Numerical tests
performed with {\sc Superbox} show that for high particle
numbers, $H_{000} = 4/3$ yields a slightly slower drift in total
energy with time (see also Section~\ref{sec:conserv}) than
$H_{000} = 1$ or 1.5, while for low particle numbers (compared to
the number of grid-cells) $H_{000} = 1$ is the better choice.
The reason for this is that for high particle numbers the
collective force dominates over the particle--particle (pairwise)
force.  On the other hand, for low particle numbers (only one or
a few particles are together in the same cell) pairwise forces
are more dominant, and it is better to exclude 'self-gravity'
(i.e. a particle feels its own force inside its cell), with
$H_{000} = 1$.  This is consistent with analytical arguments by
D.~Pfenniger (private communication).  He shows that in one
dimension, $H_{0} = \ln 4$ minimises the power at the grid
wave-number $n/2$ for the kernel (compare with Eq.\ \ref{f2.0c}) 
\begin{eqnarray}
   H_{i} & = & 1/|i|\ ,\ \ \  0 \ < \ \left|i\right| \ \leq \
   n/2, \\ 
   H_{i} & = & \ln 4\ ,\ \ \  i\ = \ 0. \nonumber
\end{eqnarray}
He finds that the power is exactly cancelled at the wavenumber
$n/2$, and for finite $n$ can also be cancelled for a value of
$H_{0}$ close to $\ln 4$.  For $H_{0} > \ln 4$ the power at high
frequencies increases, while for $H_{0} < \ln 4$ some wavenumbers
$<n/2$ are filtered out.  In this sense, $H_{0} = \ln 4 \approx
4/3$ optimises the kernel.  However, the 3D-problem is slightly
different since forces instead of the potential are used.  The
artificial wavenumbers induced by the grid, that are to be
minimised, are then much more numerous.  This problem would need
further addressing. \\ 
The FFT-algorithm gives the exact solution of the grid-based
potential for a periodical system.  For the exact solution of an
isolated system, which is what we are interested in, the size of
the density-array has to be doubled ($2N$), filling all inactive
grid cells with zero density, and extending the Green's function
in the empty regions in the following way:
\begin{eqnarray}
  \label{eq:2.1a}
  H_{2n-i,j,k} & = & H_{2n-i,2n-j,k} \ = \ H_{2n-i,j,2n-k} \ = \ 
  H_{2n-i,2n-j,2n-k} \\
  & = & H_{i,2n-j,k} \ = \ H_{i,2n-j,2n-k} \ = \ H_{i,j,2n-k} \ =
  \ H_{i,j,k}. \nonumber 
\end{eqnarray}
This provides the {\it isolated solution} of the potential in the
simulated area between $i,j,k=0$ and $N-1$.  In the inactive part
the results are unphysical.  To keep storage as low as possible,
only a $2N \times 2N \times N$-array is used for transforming the
densities, and a $(N+1) \times (N+1) \times (N+1)$-array is used
for the Green's function.  For a detailed discussion see Eastwood
\& Brownrigg (1978) and also Hockney \& Eastwood (1981). \\
The FFT-routine incorporated in {\sc Superbox} is a simple
one-dimensional FFT and is taken from Werner \& Schabach (1979)
and Press et al.\ (1986).  It is fast and makes the code portable
and not machine specific.  The low-storage algorithm to extend
the FFT to 3 dimensions, to obtain the 3-D-potential, is taken
from Hohl (1970).  The performance of {\sc Superbox} can be
increased by incorporating machine-optimised FFT routines.

\subsection{Acceleration and orbit integration}
\label{sec:accel}

\noindent
In this section we explain the force calculation performed in
{\sc Superbox}, and compare it with standard methods. We follow
closely the notation of Hockney \& Eastwood (1981) and Couchman
(1999) to document the differences. 

With the mass density as an example we show, in a formal
way, how its values on a 3D mesh are obtained. Using
the $\delta$-functional for the particle shape function,
a particle distribution $n({\bf x})$ is given by
\begin{eqnarray}
  n({\bf x}) & = & \sum_{\alpha} \delta^{3}({\bf x} - {\bf
  x}_{\alpha})\ , 
\end{eqnarray}
where $\alpha$ is a summation index running over all particles,
and ${\bf x}_\alpha=(x_{\alpha},y_{\alpha},z_{\alpha})$ is the
position vector of particle $\alpha$. As smoothing kernel
we select here
\begin{eqnarray}
  W({\bf x}, {\bf\Delta x}) & = & \Pi\Bigl({x\over\Delta x}\Bigr)
                   \cdot \Pi\Bigl({y\over\Delta y}\Bigr) \cdot
                   \Pi\Bigl({z\over\Delta z}\Bigr) \ .\nonumber
\end{eqnarray}
${\bf\Delta x}$ denotes a vector of smoothing lengths
$(\Delta x,\Delta y,\Delta z)$ in three coordinate
directions. $W$ is a three dimensional generalisation of
the standard top-hat function
\begin{eqnarray}
  \Pi(\xi) & = & \cases{\displaystyle 0 \ , & $\vert \xi \vert >
    {1\over 2}$,  \cr 
    {1\over 2} \ , & $\vert \xi \vert = {1\over 2}$, \cr 
    1        \ , & $\vert \xi \vert < {1\over 2}$ \
               . \cr }  \nonumber
\end{eqnarray}
Hence the smoothed mass-density field is
\begin{eqnarray}
  \varrho({\bf x}) & = & m \cdot \, W \star n
  \ = \ m \cdot \int W({\bf x}-{\bf x}^{\prime}, {\bf\Delta x}) \,
  n({\bf x}^{\prime}) \ {\rm d}^3{\bf x}^{\prime} \ , 
\end{eqnarray}
where $m$ denotes the particle mass,
and the $\star$-operator a convolution as defined by the
equation above. 
Mathematically we get from $\varrho({\bf x})$
a so-called mesh sampled functional $\dbg{\varrho}({\bf x})$
on a three dimensional mesh by
\begin{eqnarray}
  \dbg{\varrho}({\bf x}) & = & \comb({\bf x}) \cdot \varrho({\bf
  x}), 
\end{eqnarray}
using the three-dimensional comb or sampling function defined as
\begin{eqnarray}
  \comb({\bf x}) & = & \sum_{i,j,k=0}^N \delta(x-x_{ijk})
  \cdot\delta(y-y_{ijk})\cdot\delta(z-z_{ijk})\ , \nonumber
\end{eqnarray}
where $i,j,k$ are indices of grid cells, whose centres are 
located at the vector points ${\bf x}_{ijk} = \{ x_{ijk},
  y_{ijk}, z_{ijk}\} $. From the functional $\dbg{\varrho}({\bf
  x})$ a finite set of discrete mesh-sampled density values
$\mesh{\varrho} = \{\varrho_{ijk}, i,j,k=0,1,...,N\}$ is
obtained, which are defined at the centres of the grid cells.  For
each cell centre or mesh point they are computed by integrating
$\dbg{\varrho}$ over an arbitrarily shaped and sized volume
containing just this and only this mesh point.  The standard
discrete FFT procedure (see above) is applied to the set of
numbers $\mesh{\varrho}$ in order to determine the grid-based
potential $\mesh{\Phi}=\{\Phi_{ijk}, i,j,k=0,1,...,N\}$.
$\mesh{\Phi}$ is a set of numbers resulting from the FFT
and represents the mesh sampled functional
$\dbg{\Phi}({\bf x})=\comb({\bf x})\Phi({\bf x})$,
where $\Phi({\bf x})$ is a true smooth gravitational
potential. The set of values $\mesh{\Phi}$ is obtained from 
$\dbg{\Phi}$ by integrating over volumes containing single mesh
points, just in the same way as explained above for $\varrho$.
We define the one-dimensional two-point difference operator for
$x$ direction in the standard way (Hockney \& Eastwood 1981,
p. 163),  
\begin{eqnarray}
  D_x(x,y,z,\Delta x) & = & {1 \over 2 \Delta x}\ \Bigl(\delta(x +
  \Delta x) - \delta(x - \Delta x)\Bigr)\delta(y)\delta(z) \ . \nonumber
\end{eqnarray} 
To keep the notation clearer we remain for the moment in one
dimension. A first order difference approximation to the
x-component of the acceleration vector is given by
\begin{eqnarray}
 a_x\one(x,y,z,\Delta x) & = & D_x \star \Phi(x) = 
 {\Phi(x+\Delta x,y,z) - \Phi(x-\Delta x,y,z) \over 2\Delta x}
        \ .
\end{eqnarray}
Sampling this on the mesh yields
\begin{eqnarray}
  \dbgx{a}\one(x,y,z,\Delta x) & = & \comb({\bf x}) \cdot
  a_x\one(x,y,z) \nonumber \\
  & = & \comb({\bf x}) \cdot {\Phi(x+\Delta x,y,z) -
  \Phi(x-\Delta x,y,z)  \over 2\Delta x}
 \ .
\label{eqn:acce}
\end{eqnarray}
If this is integrated over volumes containing a single
mesh point, again exactly as described above for
the cases of $\varrho$ and $\Phi$, a set
of accelerations $a_{ijk}\one$ defined at the mesh point
centres results:
\begin{eqnarray}
  a_{x,ijk}\one & = & {\Phi_{i+1,jk} - \Phi_{i-1,jk}\over 2\Delta
  x} \ ; 
\end{eqnarray}
here we have assumed that $2 \Delta x = x_{i+1,jk}-x_{i-1,jk}$
is selected so as to match with the mesh point distances.  In our
notation the set of values $\{a_{x,ijk}\one,
i,j,k=0,1,\ldots,N\}$ is denoted by $\meshx{a}\one$.  The
corresponding vector components for the acceleration in $y$ and
$z$ direction are analogous.  In the standard particle-mesh
schemes, the accelerations between the mesh centres are obtained
by using the same smoothing kernel as for the density, i.e.
\begin{eqnarray}
  a_x\one(x,y,z,\Delta x) = W\star\dbgx{a}\one(x,y,z) = 
  a_{x,ijk}\one\ \ \ \ & {\rm for} & \ \ x-x_{ijk}<{\Delta x\over
  2} \nonumber\\ 
    & {\rm and} & \ \ y-y_{ijk}<{\Delta y\over 2} \nonumber\\
    & {\rm and} & \ \ z-z_{ijk}<{\Delta z\over 2} \ .\nonumber
\end{eqnarray}
The use of the $\delta$-functional for the particle shape and the
top-hat function for smoothing gives the rather crude standard
NGP scheme, where the acceleration on a particle is constant in
each cell and has a discontinuity at the cell boundaries.  A
possible improvement regarding energy and momentum conservation
is to use a so-called cloud-in-cell or CIC scheme, where the
shape function of particles is a top-hat function, and the
assignment function is a triangle function (see Hockney \&
Eastwood 1981).  However, as one can also see in the cited book, 
such schemes suffer from force anisotropies, despite having very
good conservation properties (i.e. the magnitude and direction of
the force error depends on whether one goes parallel to the mesh
coordinates or not). \\
Therefore a non-standard scheme is applied in {\sc Superbox}.  It
is NGP in nature, provides a very good energy and angular
momentum conservation and a significant improvement for force
anisotropies.  Note that the rather involved mathematical
formalism above will later yield significant hindsight what is
special with {\sc Superbox}.  The notation follows closely
Hockney \& Eastwood's and Couchman's. \\  
We begin with a Taylor expansion of the acceleration
(for example the $x$-compo\-nent) centred on $(x,y,z)$,
the position of the centre of cell $i,j,k$:
\begin{eqnarray}
  a_x\two(x+dx,y+dy,z+dz) & = & a_x\two(x,y,z) \ + \nonumber \\ 
  \frac{\partial a_x}{\partial x}(x,y,z) \cdot dx & + &
  \frac{\partial a_x}{\partial y}(x,y,z) \cdot dy \ + \
  \frac{\partial a_x}{\partial z}(x,y,z) \cdot dz \\
    & + & {\mathcal O} ({\bf dx}^{2}), \nonumber
\end{eqnarray}
where ${\mathcal O}({\bf dx}^{2})$ denotes any higher order
terms in $dx$, $dy$, $dz$ or combinations thereof. 
The displacements $dx$, $dy$ and $dz$ should not be
linked or confused with the mesh spacings
$\Delta x$, $\Delta y$, $\Delta z$. They are free
and should provide an interpolated expression for
$a_x$ at any point inside a given mesh cell, e.g.
a particle's position. We then use the difference
approximations
\begin{eqnarray}
  a_x\two(x,y,z,\Delta x) & = & 
  \frac{\partial \Phi}{\partial x}(x,y,z,\Delta x)
    = D_x\star\Phi \ , \nonumber \\
  \frac{\partial a_x}{\partial x}(x,y,z,\Delta x,\Delta y,\Delta
  z) & = & \frac{\partial^{2} \Phi}{\partial x^{2}}(x,y,z,\Delta
  x,\Delta y,\Delta z) \equiv D_{xx}\star\Phi \ , \nonumber \\
  \frac{\partial a_x}{\partial y} (x,y,z,\Delta x,\Delta y,\Delta
  z) & = & \frac{\partial^{2} \Phi}{\partial x \partial y}
  (x,y,z,\Delta x,\Delta y,\Delta z) \equiv D_{xy}\star\Phi \ ,
  \nonumber \\  
  \frac{\partial a_x}{\partial z} (x,y,z,\Delta x,\Delta y,\Delta
  z) & = & \frac{\partial^{2} \Phi}{\partial x \partial z}
  (x,y,z,\Delta x,\Delta y,\Delta z) \equiv D_{xz}\star\Phi \ . 
\end{eqnarray}\\
We may now generate a mesh sampled $\dbg{a_x}\two = \comb a_x\two
$ in the standard way and integrate this over any volume
containing one grid cell centre to obtain a set of acceleration
values $\meshx{a}\two$ given by the following expressions:
\begin{eqnarray}
  \label{eq:2.2}
  a_{ijk,x}\two(dx,dy,dz) & = & \frac{\Phi_{i+1,jk} -
  \Phi_{i-1,jk}}{2 \Delta x}   \\ 
  & + & \frac{\Phi_{i+1,jk} + \Phi_{i-1,jk} - 2 \cdot
    \Phi_{ijk}} {(\Delta x)^{2}} \cdot dx \nonumber \\ 
  & + & \frac{\Phi_{i\!+\!1,j\!+\!1,k} - \Phi_{\!i-\!1,j\!+\!1,k}
    + \Phi_{i\!-\!1,j\!-\!1,k} - \Phi_{i\!+\!1,j\!-\!1,k}} {4
    \Delta x \Delta y} \cdot dy \nonumber \\ 
  & + & \frac{\Phi_{i\!+\!1,j,k\!+\!1} - \Phi_{i\!-\!1,j,k\!+\!1}
    + \Phi_{i\!-\!1,j,k\!-\!1} - \Phi_{i\!+\!1,j,k\!-\!1}} {4
    \Delta x \Delta z} \cdot dz \nonumber 
\end{eqnarray}
Note that we generally take $\Delta x=\Delta y=\Delta z = 1$
i.e.\ the cell-length is assumed to be equal along the three axes
and unity.  The accelerations in $y$- and $z$-direction are
calculated analogously. \\
At first glance the interpolation scheme Eq.~\ref{eq:2.2} for the
acceleration seems to be closely related to the procedure used in
a standard cloud-in-cell (CIC) scheme.  In one dimension, CIC and
{\sc Superbox} methods can be more easily compared to each other.
For $0 \leq dx < \Delta x/2$ one has 
\vspace*{-0.1cm}
\begin{eqnarray}
  \label{eq:ngpcic}
 {\rm CIC}\ : \ a(x + dx) & = &  a_{i} \cdot \frac{\Delta x -
  dx}{\Delta x} \ + \ a_{i+1} \cdot \frac{dx}{\Delta x} \nonumber
  \\ 
  & = & \frac{\Phi_{i+1} - \Phi_{i-1}}{2 \Delta x} \ + \ \frac{
  \Phi_{i+2} - \Phi_{i+1} - \Phi_{i} + \Phi_{i-1}}{2 (\Delta
  x)^{2}} \ dx\ , \nonumber \\  
 \mathsf{Superbox}\ : \ a(x + dx) & = & a_{i}\ + \ \left. \frac{
  {\rm d}a} {{\rm d}x} \right|_{i} \cdot dx \nonumber \\
  & = & \frac{\Phi_{i+1} - \Phi_{i-1}}{2 \Delta x} \ + \ \frac{
  \Phi_{i+1} + \Phi_{i-1} - 2 \Phi_{i}} {(\Delta x)^{2}}\ dx\ ,
  \nonumber 
\end{eqnarray}
where $i$ denotes the cell-index. Clearly this is different,
because CIC uses accelerations in {\em two} neighbouring cells,
while {\sc Superbox} takes acceleration and its derivative only
in {\em one} cell.  Therefore we propose to consider {\sc
  Superbox} as an NGP type scheme, although a second order
interpolated acceleration is used, because nowhere the
acceleration of the neighbouring cell is entering into the
equations.  Though of NGP nature, $a_x$ in {\sc Superbox} behaves
steadily when crossing cell boundaries along the $x$-axis (and
$a_y$, $a_z$ correspondingly on their respective axes).  In
contrast to CIC our scheme is not steady in any other direction.
However, the errors induced by that are much smaller than in a
standard NGP scheme due to the above higher order interpolation, 
and are perfectly tolerable as will be shown throughout this
paper. 

\begin{figure}[h!]
  \begin{center}
    \leavevmode
    \epsfxsize=12cm
    \epsfysize=8.5cm
    \epsffile{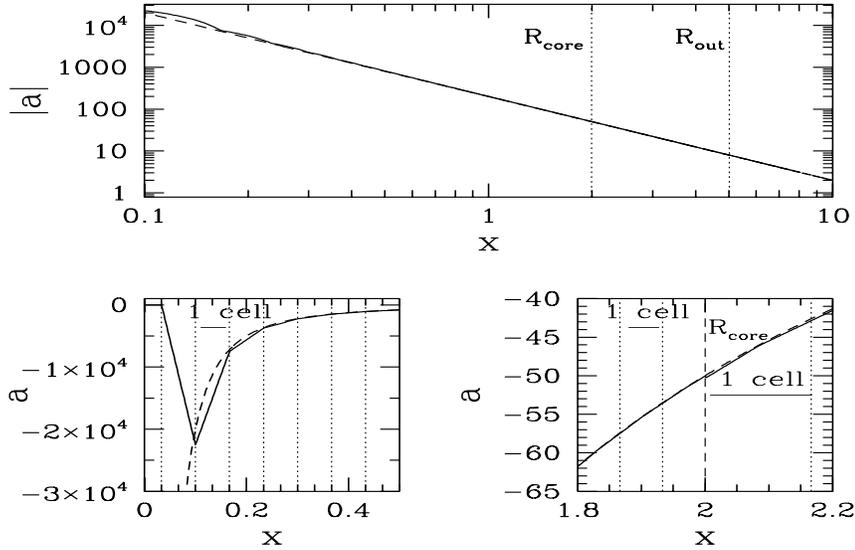}
    \caption{ 
      Pairwise acceleration calculated by {\sc Superbox} (using
      $H_{000} = 1$) for one particle fixed in the centre and the
      second moving along the $x$-axis. The agreement between the
      acceleration calculated by {\sc Superbox} (solid line) and
      the analytical (i.e. Keplerian) force (dashed line) is very
      good, if the two particles are at least 2 cells apart.  Upper
      panel: pairwise acceleration over the entire spatial range.
      Lower panel: blown up pictures for the innermost part
      (left) and the grid-boundary (right). }
    \label{forcex}
  \end{center}
\end{figure}

\begin{figure}[h!]
  \begin{center}
    \leavevmode
    \epsfxsize=12cm
    \epsfysize=8.5cm
    \epsffile{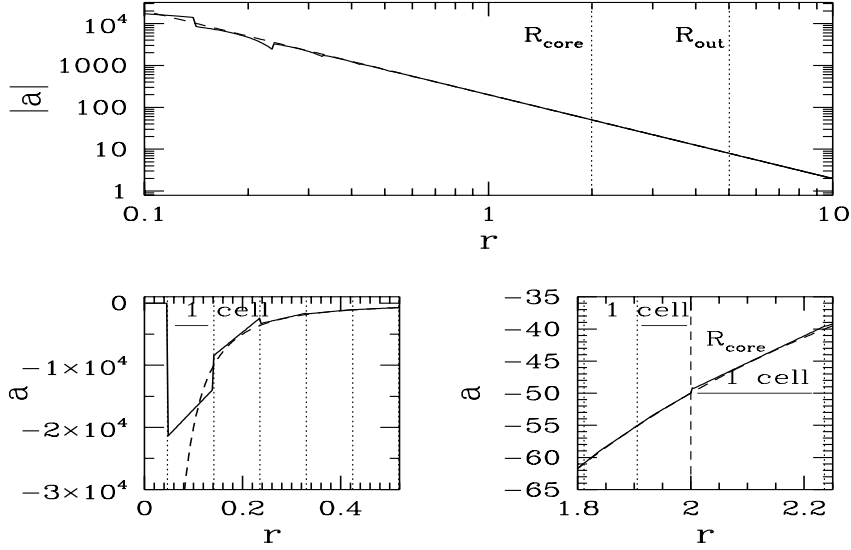}
    \caption{As Fig.~\ref{forcex}, except that here the second
      particle moves along a diagonal  with $y = x \ , \ z=0 \ ,
      \ r = \sqrt{x^{2}+y^{2}}$. }
    \label{forcexy}
  \end{center}
\end{figure}

The salient feature of {\sc Superbox} regarding the force
anisotropies can be understood by considering the force at a
distance $r = \left| \mathbf{dr}\right|$ from a particle located
at the centre of a cell.  To lowest order the {\sc Superbox}
force-error $\varepsilon$ is then
\begin{eqnarray}
  \label{isoerr}
  \vec{\varepsilon} & = & - \frac{{\bf dr}}{r^{3}} \ ,
\end{eqnarray}
where ${\bf dr} = (dx,dy,dz)$ is the displacement vector from the
cell centre, and we have neglected all terms of second order in
$\Delta x$, $\Delta y$, or $\Delta z$ in a Taylor series
expansion of the potential differences.  In other words to lowest
order the {\sc Superbox} force error points to the exact centre
of the cell.  Thus, compared with standard NGP schemes, we get a
very precise force-calculation in {\sc Superbox}, as shown in
Figs.~\ref{forcex} and \ref{forcexy}. \\ 

The orbits of the particles are integrated forward in time using
the leapfrog scheme.  For example, for the $x$-components of the
velocity, $v_{x}$, and position, $x$, vectors of particle~$l$,
\begin{eqnarray}
  v_{x,l}^{n+1/2} & = & v_{x,l}^{n-1/2} + a_{x,l}^{n} 
   \cdot \Delta t \label{f2.2a} \\ 
  x_l^{n+1}   & = & x_l^{n} +  v_{x,l}^{n+1/2}  \cdot \Delta t,
   \nonumber 
\end{eqnarray}
where $n$ denotes the $n$th time-step and $\Delta t$ is the
length of the integration step. \\

\subsection{The grids}
\label{sec:grid}

\begin{figure}[h!]
  \begin{center}
    \leavevmode
    \epsfxsize=8.0cm
    \epsfysize=8.cm
    \epsffile{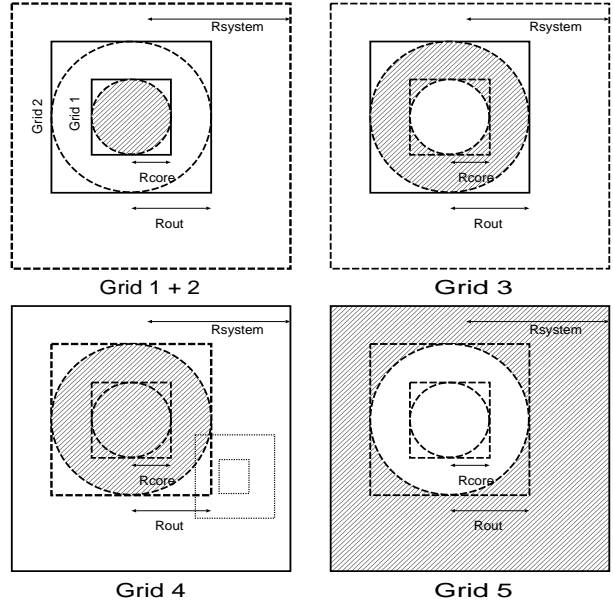}
    \caption{
    The five grids of {\sc Superbox}.  In each panel, solid lines
    highlight the relevant grid.  Particles are counted in the
    shaded areas of the grids.  The lengths of the arrows are
    $(N/2) - 2$ grid-cells (see text).  In the bottom left panel,
    the grids of a hypothetical second galaxy are also shown as
    dotted lines. } 
  \label{gitter}
  \end{center}
\end{figure}

\noindent
For each galaxy, 5~grids with~3 different resolutions are used.
This is made possible by invoking the additivity of the potential
(Fig.~\ref{gitter}). 

\noindent
The five grids are as follows:
\begin{itemize}
\item Grid 1 is the high-resolution grid which resolves the
  centre of the galaxy.  It has a length of $2 \times R_{\rm
  core}$ in one dimension.  In evaluating the densities, all
  particles of the galaxy within $r \leq R_{\rm core}$ are stored
  in this grid. 
\item Grid 2 has an intermediate resolution to resolve the galaxy
  as a whole.  The length is $2 \times R_{\rm out}$, but only
  particles with $r \leq R_{\rm core}$ are stored here, i.e.\
  the same particles as are also stored in grid~1.
\item Grid~3 has the same size and resolution as~grid 2, but it
  only contains particles with $R_{\rm core} < r \leq R_{\rm
  out}$.  
\item Grid~4 has the size of the whole simulation area (i.e.\
  'local universe' with $2 \times R_{\rm system}$), and has the
  lowest resolution.  It is fixed.  Only particles of the galaxy
  with $r \leq R_{\rm out}$ are stored in grid 4. 
\item Grid~5 has the same size and resolution as grid~4.  This
  grid treats the escaping particles of a galaxy, and contains
  all particles with $r > R_{\rm out}$.
\end{itemize}
Grids~1 to~3 are focused on a common centre of the galaxy and
move with it through the 'local universe', as detailed below.
All grids have the same number of cells per dimension, $N$, for
all galaxies. The boundary condition, requiring two empty cells
with $\varrho = 0$ at each boundary, is open and non-periodic,
thus providing an isolated system. This however means that only
$N-4$ active cells per dimension are used. \\
To keep the storage requirement low, all galaxies are treated
consecutively in the same grid-arrays, whereby the particles
belonging to different galaxies can have different masses.  Each
of the~5 grids has its associated potential $\Phi_{\rm i},\ i =
1,2,...,5$ computed by the PM technique from the particles of one
galaxy located as described above.  The accelerations are
obtained additively from the~5 potentials of each galaxy in turn
in the following way:
\vspace*{-0.2cm} 
\begin{eqnarray}
  \label{eq:addphi}
  \Phi(r) & = & \left[ \theta(R_{\rm core}-r) \cdot \Phi_{1} +
  \theta(r-R_{\rm core}) \cdot \Phi_{2} + \Phi_{3} \right] \cdot
  \theta(R_{\rm out}-r) \\
  & + & \theta(r-R_{\rm out}) \cdot \Phi_{4} + \Phi_{5}\ ,
  \nonumber \\
  \Phi(R_{\rm core}) & = & \Phi_{1} + \Phi_{3} + \Phi_{5} \nonumber
  \\
  \Phi(R_{\rm out}) & = & \Phi_{2} + \Phi_{3} + \Phi_{5} \nonumber
\end{eqnarray}
where $\theta(\xi)=1$ for $\xi > 0$ and $\theta(\xi)=0$ otherwise.
This means: 
\begin{itemize}
\item For a particle in the range $r \leq R_{\rm core}$, the
  potentials of grids 1, 3 and 5 are used to calculate the
  acceleration. 
\item For a particle with $R_{\rm core} < r \leq R_{\rm out}$,
  the potentials of grids~2, 3~and 5~are combined.
\item And finally, if $r > R_{\rm out}$ then the acceleration is
  calculated from the potentials of grids~4 and~5.
\item A particle with $r > R_{\rm system}$ is removed from the
  computation.
\end{itemize}
Due to the additivity of the potential (and hence its derivatives,
the accelerations) the velocity changes originating from the potentials
of each of the galaxies can be separately updated and
accumulated in the first of the leap-frog formulas
Eq.~\ref{f2.2a}. The final result does not depend on the order by which
the galaxies are taken into account and it could be computed even in
parallel, if a final accumulation takes place. After all
velocity changes have been applied in all galaxies,
the positions  
of the particles are finally updated (see Fig.~\ref{super}). \\
As long as the galaxies are well separated, they feel only the
low-resolution potentials of the outer grids.  But as the galaxies
approach each other their high-resolution grids overlap, leading
to a high-resolution force calculation during the interaction.

\subsubsection{Grid tracking}
\label{sec:tracking}

\noindent
Two alternative schemes to position and track the inner and
middle grids can be used.  The most useful scheme centres the
grids on the density maximum of each galaxy at each step.  The
position of the density maximum is found by constructing a sphere
of neighbours centred on the densest region, in which the centre
of mass is computed. This is performed iteratively.  The other
option is to centre the grids during run-time on the position of
the centre of mass of each galaxy using all its particles
remaining in the computation.  

\subsubsection{Edge-effects}
\label{sec:edge}

\begin{figure}[!h]
  \begin{center}
    \leavevmode
    \epsfxsize=10cm
    \epsfysize=12cm
    \epsffile{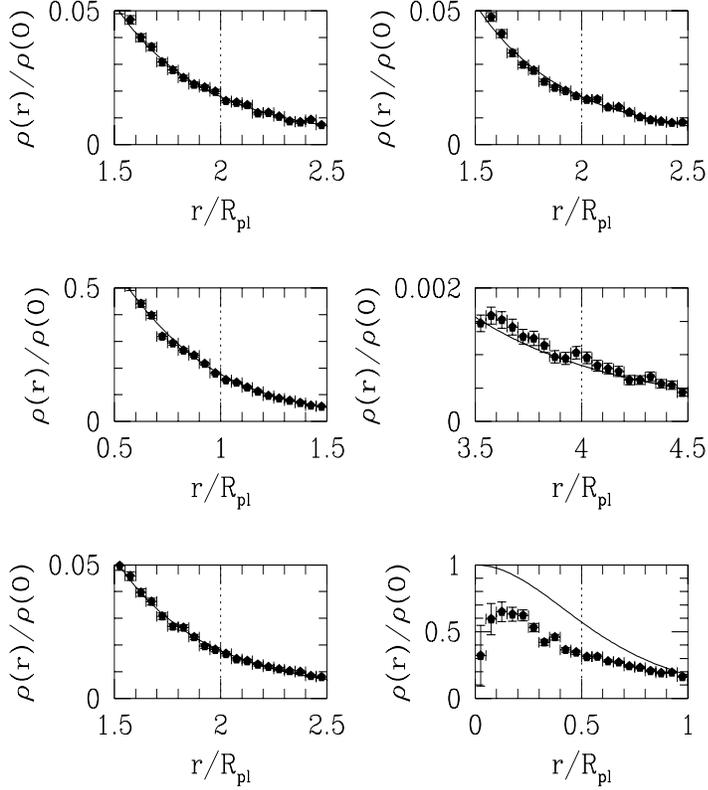}
    \caption{
      The radial density distribution of Plummer spheres at the
      grid-boundary between the innermost and the middle grid
      (shown as dotted vertical lines) after computing for
      10~half-mass crossing times for different grid-sizes.  The
      radial density profile resulting from the {\sc Superbox}
      computation, always with $N_{\rm p}=5\times10^4$, is shown
      as dots with error-bars.  The analytical density law is
      shown as the solid line.  In all cases except the bottom
      left panel $R_{\rm out}=9R_{\rm pl}$ (where $R_{\rm pl}$
      is the Plummer radius).
      Upper left: $N=32$ grids per dimension, $R_{\rm core} =
      2R_{\rm pl}$.  Upper right: $N=64$, $R_{\rm core} = 2R_{\rm
      pl}$.  Middle left: $N=32$, $R_{\rm core} = 1R_{\rm pl}$.
      Middle right: $N=32$, $R_{\rm core} = 4R_{\rm pl}$.  Bottom
      left: $N=32$, $R_{\rm core} = 2R_{\rm pl}$ and $R_{\rm out}
      = 14R_{\rm pl}$.  Bottom right: $N=32$, $R_{\rm core} = 0.5
      R_{\rm pl}$.  Only if the innermost grid is chosen too
      small (bottom right) are there significant deviations from
      the analytical profile. For this calculations $H_{000}=4/3$
      is used.} 
    \label{edge}
  \end{center}
\end{figure}

\noindent
It is shown in Fig.~\ref{gitter} that only spherical regions of the
cubic grids contain particles (except for Grid 5).  Particles with
eccentric orbits can cross the border of two grids, thus being
subject to forces resolved differently.  No interpolation of the
forces is done at the grid-boundaries.  This keeps the code fast
and slim, but the grid-sizes have to be chosen properly in
advance to minimise the boundary discontinuities.  As shown in
Section~\ref{sec:relax}, this leads to some additional but
negligible relaxation-effects, because the derived total
potential has insignificant discontinuities at the
grid-boundaries (Wassmer 1992).  The best way to avoid these
edge-effects is to place the grid-boundaries at 'places' where
the slope of the potential is not steep. \\
Fig.~\ref{edge} shows that, as long as the grid-sizes
are chosen properly, there are no serious effects at the
grid-boundaries.  Only in the example shown in the bottom-right
panel the innermost grid was chosen too small, leading to a low
particle number per grid-cell.  Performing such computations with
$H_{000}=4/3$ leads to spurious results, as explained in
Section~\ref{sec:green}.  Nevertheless the profile is smooth at the
grid-boundary. 

\subsubsection{Model units}
\label{sec:mu}

\noindent 
{\sc Superbox} employs model units, with the gravitational
constant in model units being $G_{\rm mod} = 2$ owing to
historical reasons.  A flag specifies if the user wants to input
and output data in physical units (i.e. kpc, M$_{\odot}$, km/s,
Myr).  Scaling to and from physical units is achieved via the
formulae: 
\begin{eqnarray}
  \left( {L_{\rm phys} \over L_{\rm mod}} \right)^3 & = & {G_{\rm 
      phys} \over G_{\rm mod}} \, {M_{\rm phys} \over M_{\rm
      mod}} \, \left( {T_{\rm phys} \over T_{\rm mod}} \right)^2,
  \label{eq:scaling} \\
  {V_{\rm phys} \over V_{\rm mod}} & = & {L_{\rm phys}\over
    T_{\rm phys}} {T_{\rm mod} \over L_{\rm mod}}, \nonumber
\end{eqnarray}
where $L$ is the length, $M$ the mass, $T$ the time, $V$ the
velocity and $G$ the gravitational constant.  The model length
unit is taken to be $L_{\rm mod}=1$ for the length of $R_{\rm
  core,1}$ of galaxy~1.  The corresponding physical length scale, 
$L_{\rm phys}$, is taken from the input data that specifies the
length of this grid.  The physical mass of the first galaxy,
$M_{\rm phys,1}$, is taken to be unity in model units, $M_{\rm
  mod,1 }=1$.  The integration step-size, $T_{\rm phys}$, is
specified by the user.  The step-size in model units, $T_{\rm
  mod}$, follows from Eq.~\ref{eq:scaling}. \\
Converting factors for each grid of all galaxies are calculated
to get the forces and accelerations, due to the FFT and the
differentiation of the potentials requiring a cell-length of
unity. 


\section{Conservation of Energy and Angular Momentum}
\label{sec:conserv}

\noindent
In order to check how well energy and angular momentum are
conserved, calculations with isolated Plummer-spheres in virial
equilibrium with different particle numbers, time-steps and
grid-resolutions are presented here.  To have a non-vanishing
angular momentum vector in $z$-direction for the isolated galaxy
model, all particle orbits are taken to have the same direction
in the $xy$-plane.  This procedure does not change any other 
properties of the Plummer-model, but makes it possible to check
for relative changes in angular momentum.

\subsection{Total Energy}
\label{sec:is_energy}

\noindent
Calculating the correct potential energy of a stellar system
requires summing all two-body interactions between all particles.
This is quite impossible for $N_{\rm p}\simgreat10^6$.  By using
the available grid-based approximation and adding up only the
mean potential values of the cells for each particle, a useful
estimate of the total potential energy is obtained.  If, on the
other hand, the deviation from the centre of the cell and
contributions from the neighbouring cells were taken into account
to obtain a more accurate estimate, then CPU-time would have to
be used for an arguably purposeless endeavour.  For simplicity,
in {\sc Superbox} the total potential energy is computed
without any interpolation inside grid cells (as it was used for
the acceleration); only the potential value of the
particle's cells are taken into account to compute the potential
energy.  The total potential energy at the $n$th time-step is 
thus  
\begin{eqnarray}
    E_{\rm pot}^{n} & \approx & - \frac{1}{2} \sum_{i=1}^{N_{\rm
    p}} m_{i} \Phi^{n}(x_{i},y_{i},z_{i}), \label{energy_1}
\end{eqnarray}
where $\Phi^n$ is the sum of the potentials of all grids and
galaxies used to calculate the force on particle $i$ at the
$n$-th time-step, or in other words, all potentials which
contribute to the leap-frog update of the velocity in Eq.~\ref{f2.2a}.
$m_{i}$ is the mass of the particle (equal
for all particles belonging to the same galaxy). 

The total kinetic energy is also calculated only approximately
during runtime.  This comes about because in the leapfrog
integration-scheme the velocities are half a time-step behind the 
positions.  Within our integration scheme the time-centred
interpolation for the kinetic energy would be
\begin{eqnarray}
  E_{\rm kin}^n & = & \frac{1}{2} \ \sum_{i=1}^{N_{\rm p}} m_i
  \sum_{j=1}^3 \left( \frac{v_{i,j}^{n+1/2} + v_{i,j}^{n-1/2}}
  2 \right)^2, \label{energy_2}
\end{eqnarray}
where $v^{n}_{i,j}$ is the $j$-component of the velocity vector
of particle $i$ at time-step $n$. In {\sc Superbox} the
philosophy is to keep the amount of storage as small as possible,
so that only the current velocities are stored.  To keep the
error in $E_{\rm kin}$ as small as possible, particle kinetic
energies are calculated before and after the velocity vectors are
updated in one time-step.  From these two values the arithmetic
average is taken: 
\begin{eqnarray}
  E_{\rm kin}^n \approx \frac{1}{4} \ \sum_{i=1}^{N_{\rm p}} m_{i}
  \sum_{j=1}^{3} \left( \ (v_{i,j}^{n-1/2})^{2} +
  (v_{i,j}^{n+1/2})^{2} \right). \label{energy_3}
\end{eqnarray}
This still implies an additional error of the order of $(\Delta
v)^{2}$, but keeps storage low and the code fast. 

\begin{figure}[h!]
  \begin{center}
    \leavevmode
    \epsfxsize=8cm
    \epsfysize=8cm
    \epsffile{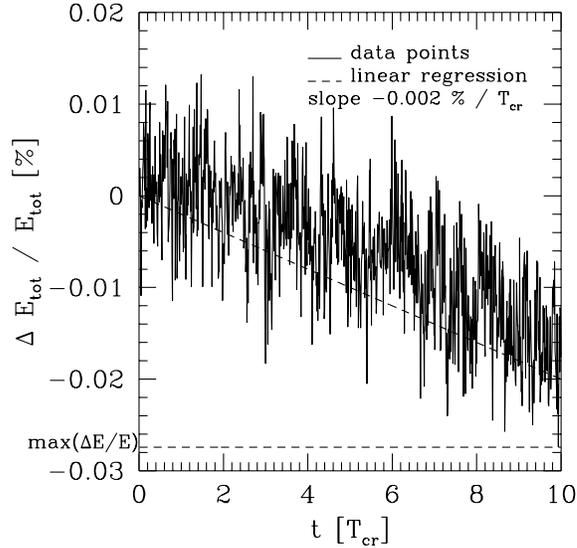}
    \caption{Linear drift of the total energy $\Delta E_{\rm
        tot}(t) / E_{\rm tot}(0)$ in per cent for $N_{\rm p}=10^6$
      particles and $N=64$ grids per dimension using $\Delta t=
      T_{\rm cr}/100$ (the regression line is shown with a
      little offset for clarity).}
    \label{run6e}
  \end{center}
\end{figure}

To quantify the performance of {\sc Superbox} in terms of
conservation of total energy, $E_{\rm tot}(t)=E_{\rm pot}(t) +
E_{\rm kin}(t)$, calculations with different time-steps, particle
number and grid resolution are done.  We consider the relative
change $\Delta E_{\rm tot}(t) / E_{\rm tot}(t_{0})$, where
$\Delta E_{\rm tot}(t) = E_{\rm tot}(t) - E_{\rm tot}(t_{0})$,
and $t_{0}$ is the reference time. It is chosen to be zero when the
adjustment to equilibrium after setting up the discrete rendition
of the galaxy is accomplished.  Max$(\Delta E_{\rm tot}(t) /
E_{\rm tot}(0))$ denotes the largest deviation in energy that
occurs during a computation (see Fig.~\ref{run6e}). 
 
Table~\ref{tab:deltat} shows that deviations in $E_{\rm tot}(t)$
are less than 0.5~per cent, as long as the time-step, $\Delta t
\simless T_{\rm cr}/10$, where $T_{\rm cr}$ is the half-mass
crossing-time of the Plummer-model.  We consider $\Delta t \leq
T_{\rm cr}/50$ to be a safe choice.  In theory (see Hockney \&
Eastwood 1981) the error of the leapfrog integrator should
decrease as $(\Delta t)^{2}$ for $\Delta t \rightarrow 0$, but as
is evident from Table~\ref{tab:deltat}, other error sources
become prominent at a sufficiently small time step. \\
\begin{table}[h!]
  \begin{center}
    \caption{Maximum deviation in total energy, max$(\Delta E_{\rm
      tot}/E_{\rm tot})$ after $10\,T_{\rm cr}$ for different
      time-steps using $N_{\rm p}=10^5$ particles and $N=32$
      grids per dimension.}
    \begin{tabular}[h!]{l|c} 
      $\Delta t$ (time-step) & max$(\Delta E_{\rm tot}(t)/E_{\rm
      tot}(0))$ \\ \hline 
      $T_{cr}/10$    & 0.4 \% \\ 
      $T_{cr}/100$   & 0.2 \% \\
      $T_{cr}/1000$ & 0.2 \% \\ 
    \end{tabular}
    \label{tab:deltat}
  \end{center}
\end{table}

Fig.~\ref{run6e} shows the time evolution of the change in total
energy for a particular calculation.  A linear drift fits the
data well, which holds true in computations extending over 50
crossing-times.  The slope of the energy-drift is 0.002~per~cent
per crossing-time.  The dispersion in energy is artificial due
to the crude calculation of potential energy.  Over a Hubble
time, which corresponds to $150\times T_{\rm cr}$ if the Plummer
model galaxy is assumed to have $T_{\rm cr}=10^8$~yr, the
resulting energy error is 0.3~per cent. \\
This linear drift is intrinsic to a PM code (see Hockney \&
Eastwood, their fig.~9.4), and its strength depends mainly on the
choice of $H_{000}$ in the Green's function.  $H_{000}$
determines the strength of the gravitational interaction of
particles in the same cell, and also the self-gravitation of a
particle (see Section~\ref{sec:green}).  Additional errors arise
through the limited grid-resolution and the approximations
adopted in evaluating the energies.  This gives an additional
oscillation around the value of the linear drift seen in
Fig.~\ref{run6e}. 
\begin{figure}[h!]
  \begin{center}
    \leavevmode
    \epsfxsize=8cm
    \epsfysize=8cm
    \epsffile{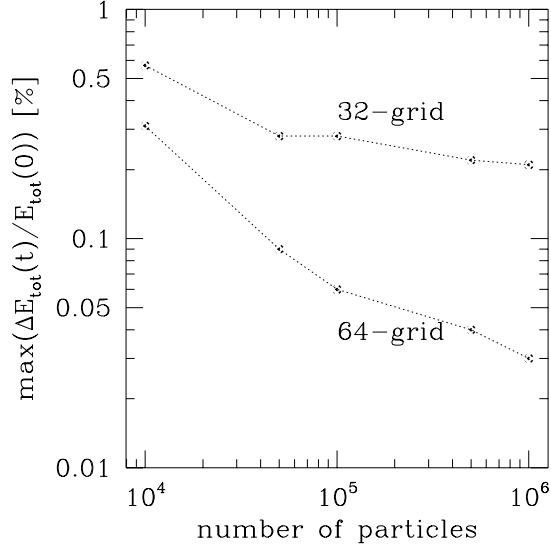}
    \caption{Maximum deviation in energy, max($\Delta E_{\rm
    tot}(t)/E_{\rm tot}(0)$) in per cent, after $10\,T_{\rm cr}$,
    for different particle numbers and  grid-resolutions, using
    $\Delta t = T_{\rm cr} / 100$.} 
    \label{is_emax}
  \end{center}
\end{figure}

Increasing the number of particles per galaxy to $N_{\rm p}>10^5$
leads to little further improvement in max$(\Delta E_{\rm tot} /
E_{\rm tot})$ with $N=32$.  The decrease of the error levels off
if the mean number of particles per cell becomes sufficiently
large.  However, changing the grid resolution from $N=32$ to 64
cells per dimension gives an improvement by a factor of at least
4, with the decrease in error still progressing for $N_{\rm p} >
10^6$ (Fig.~\ref{is_emax}).  This is due to the estimate of the
true potential by our grid-based potential, and thus of the
potential energy, improving with increasing $N$.  It also shows
that increasing the number of grid-cells while keeping the number
of particles low (with $H_{000}=4/3$) does not improve energy
conservation (e.g.\ $N=64$, $N_{\rm p}=10^{4}$).

\subsection{Angular Momentum}
\label{sec:is_angular}

\noindent
To calculate the absolute value of the total angular momentum of
the galaxy, $L_{\rm tot}$, the same technique as for the kinetic
energy is applied.  That is, the arithmetic mean of the two
velocity values (before and after the velocity update) is
calculated for each velocity component to obtain an estimate of
the velocity vector that is in-phase with the position vector. 
\begin{figure}[h!]
  \begin{center}
    \leavevmode
    \epsfxsize=8cm
    \epsfysize=8cm
    \epsffile{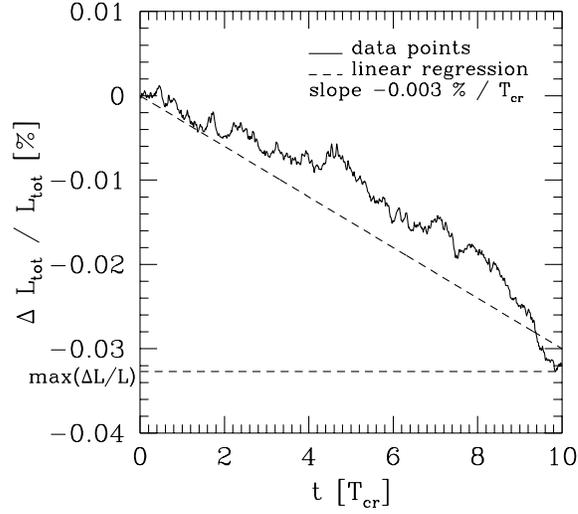}
    \caption{
      The change in $\Delta L_{\rm tot}(t) / L_{\rm
        tot}(0)$ in per~cent for $N_{\rm p}=10^6$ particles and
      $N=64$ grids per dimension using $\Delta t = T_{\rm
        cr}/100$ (the regression is shown offset).} 
    \label{run6a}
  \end{center}
\end{figure}

The time evolution of $\Delta L_{\rm tot}(t)/L_{\rm tot}(t)$ for
one particular calculation is shown in Fig.~\ref{run6a}.  As with
$E_{\rm tot}$, computations over $50\,T_{\rm cr}$ show
insignificant deviations from the linear drift.  The slope of the
drift is about 0.003~per cent per crossing-time. 

\begin{figure}[h!]
  \begin{center}
    \leavevmode
    \epsfxsize=8cm
    \epsfysize=8cm
    \epsffile{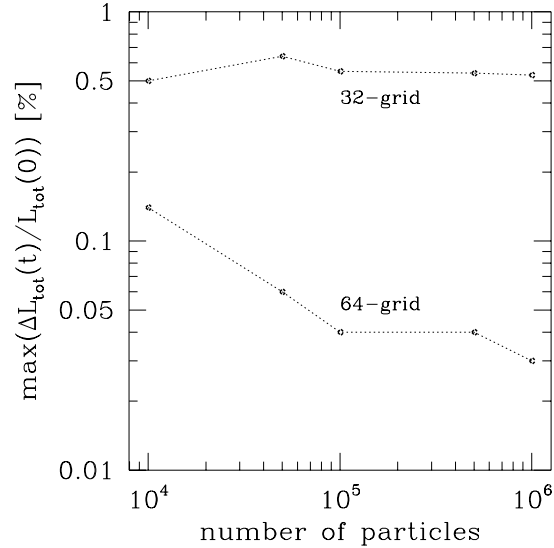}
    \caption{
      Maximum deviation in angular momentum, max$(\Delta
      L_{\rm tot}(t)/L_{\rm tot}(0))$ in~per~cent, after
      $10\,T_{\rm cr}$, for different particle numbers and
      grid-resolutions, using $\Delta t=T_{\rm cr} / 100$.
      max$(\Delta L_{\rm tot}(t)/L_{\rm tot}(0))$ is defined
      analogously to max($\Delta E_{\rm tot}(t)/E_{\rm tot}(0)$)
      in Section~\ref{sec:is_energy}.}
    \label{is_amax}
  \end{center}
\end{figure}

Conservation of $L_{\rm tot}$ does not depend on $\Delta t$ as
long as the time-step is small enough ( $\Delta t \leq T_{\rm
  cr}/50$ as for energy in Section~\ref{sec:is_energy}).  But it
is highly dependent on the grid-resolution.  Changing from $N=32$
to 64 grids per dimension improves the change in angular momentum
by at least a factor of 10.  From Fig.~\ref{is_amax} it can be
seen that max$(\Delta L_{\rm tot}/L_{\rm tot})$ is quite
independent of $N_{\rm p}$. 


\section{Relaxation}
\label{sec:relax}

\noindent
Galaxies have two-body relaxation times of the order of
$10^7$~Gyr, and show little or no relaxation over a Hubble time
even in their inner parts.  Therefore, a programme which models
galaxies has to be collision-less.  Since no particle-mesh code
can be entirely free of relaxation, we have to check on which
time-scales {\sc Superbox} provides reliable results. 

Following Standish \& Aksnes (1969), the following experiment is 
performed: A number of equal-mass and fixed particles is
distributed homogeneously inside a sphere of radius $R_{\rm sph}$
with a constant density distribution.  A second group of
particles is distributed on the surface of this sphere.  They are
allowed to move through the centre and leave the sphere on the
opposite side.  To make sure they leave the sphere, they are
given an initial non-zero radial velocity component towards the
centre of the sphere.  The points where the moving particles
leave the sphere are noted and the calculation is stopped after
all moving particles have left the sphere.  For every particle,
its deflection angle $\alpha$ is obtained from  
\begin{eqnarray}
  \cos \alpha & = & - \frac{\vec{R}_I \, \vec{R}_F}{|\vec{R}_I|
    \, |\vec{R}_F|}\;\;, 
\end{eqnarray}
\begin{figure}[h!]
  \begin{center}
    \epsfxsize=5cm
    \epsfysize=5cm
    \leavevmode
    \epsffile{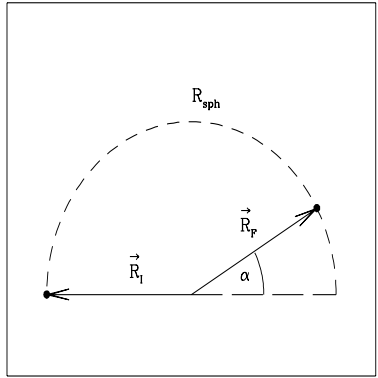}
    \caption{The deflection angle $\alpha$. }
    \label{angle}
  \end{center}
\end{figure}
where $\vec{R}_I$ and $\vec{R}_F$ are the vectors from the centre
of the sphere to the initial and final position of the particle
(see Fig.~\ref{angle}).  The mean deflection angle, ${\overline 
  \alpha}$, of all particles is computed and the relaxation time,
$T_{\rm rel}$, is calculated from 
\begin{eqnarray}
  T_{\rm rel} & = & \frac{\sin 90}{\sin {\overline
  \alpha}}\,{\overline T_{\rm cr}}.
\end{eqnarray}
Here ${\overline T_{\rm cr}}$ is the mean time the moving
particles require to travel through the sphere. \\
The number of moving particles is chosen to be $10^5$, which
gives a sufficiently accurate estimate for $T_{\rm rel}$.  The
number of fixed particles, $N_{\rm fix}$, and the size of the
inner mesh, $R_{\rm core}$, are varied.  Grids with $32$
grid-cells per dimension are used.  For a certain
combination of $N_{\rm fix}$ and $R_{\rm core}$, 100~calculations
with different random number seeds (both for the fixed and moving
particles) are performed.  In all cases, $R_{\rm sph}<R_{\rm
  out}$ (see also Fig.~\ref{gitter}). 
\begin{figure}[h!t]
  \begin{center}
    \leavevmode
    \epsfxsize=8cm
    \epsfysize=8cm
    \epsffile{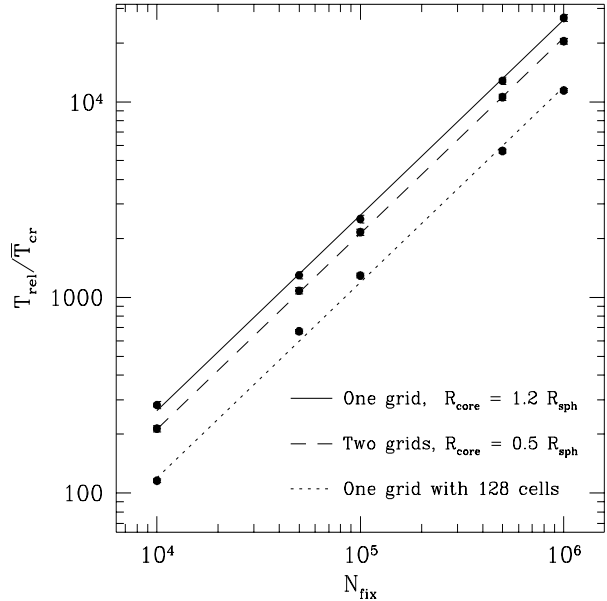}
    \caption{The ratio $T_{\rm rel}/{\overline T_{\rm cr}}$ as a 
      function of the number of fixed particles.}
    \label{trelp} 
  \end{center}
\end{figure}

Fig.~\ref{trelp} shows the ratio $T_{\rm rel}/{\overline T_{\rm
cr}}$ as a function of $N_{\rm fix}$. If one grid contains the
whole sphere ($R_{\rm core}= 1.2 \, R_{\rm sph}$) a linear
dependence of $T_{\rm rel}/{\overline T_{\rm cr}}$ on $N_{\rm
fix}$ fits the results very well.  Such a dependency is predicted
for large particle numbers by standard relaxation theory (Spitzer
\& Hart 1971; Binney \& Tremaine 1987).  For 
$N_{\rm fix} \simgreat 10^5$, $T_{\rm rel}/{\overline T_{\rm cr}}
> 10^3$.  Since $T_{\rm cr} \approx 10^8$~yr for a typical
galaxy, the relaxation time is one order of magnitude larger than
a Hubble time.  Hence galaxies can be modelled with {\sc
  Superbox} provided $N_{\rm p}>10^5$ particles per galaxy. 

These results are not substantially changed if the particles have
to cross a grid boundary.  As an example, the case $R_{\rm core}
= 0.5\,R_{\rm sph}$ is shown in Fig.~\ref{trelp}.  Compared to
the one-grid case, $T_{\rm rel}$ is reduced by about 20~per cent 
nearly independently of $N_{\rm fix}$.  Hence, grid boundaries do
not decrease $T_{\rm rel}$ significantly.  The relaxation time
also drops if the number of grid-cells is increased: Due to the
better resolution, forces from particles in adjacent cells become
larger, increasing the relaxation.  As an example the case with
128 cells per dimension is shown in Fig.~\ref{trelp}.  The
relaxation times are a factor of $2.3$ smaller compared to the
case with $N=32$.  Assuming that all particles within a cell are
located at the cell centre, and treating the deflection of the
moving particles as a pure N-body problem, we may obtain an
analytical estimate of the relaxation in {\sc Superbox}. A
consideration similar to the one described in Binney \& Tremaine
(1987) p.~187 ff shows that the
relaxation time should depend in the following way on $N_{\rm
  fix}$ and $N_{\rm gc}$:
\begin{eqnarray}
  \label{trelax}
  T_{\rm rel} & = & \kappa \cdot \frac{N_{\rm fix}} {\ln (\gamma
  N_{\rm gc})} \cdot T_{\rm cr}
\end{eqnarray}
with $\gamma = 1/3$.  From the experiments described in this
section we obtain a value for the constant of proportionality 
$\kappa = 0.05$.  Together with this value,  Eq.~\ref{trelax}
gives an estimate of the relaxation time of {\sc Superbox}.  From
this result we can quantify, for a given grid resolution and
computational time, the minimum number of particles that should
be used to exclude any unwanted relaxation effects.  It is again
obvious that calculations with a high number of grid-cells should
be done only with a sufficiently large particle number to ensure
$T_{\rm rel} > T_{\rm Hubble}$. 


\section{Memory and CPU-Time Requirements}
\label{sec:memcpu}

\subsection{Memory}
\label{sec:mem}

\begin{figure}[h!]
  \begin{center}
    \leavevmode
    \epsfxsize=14.0cm
    \epsfysize=9cm
    \epsffile{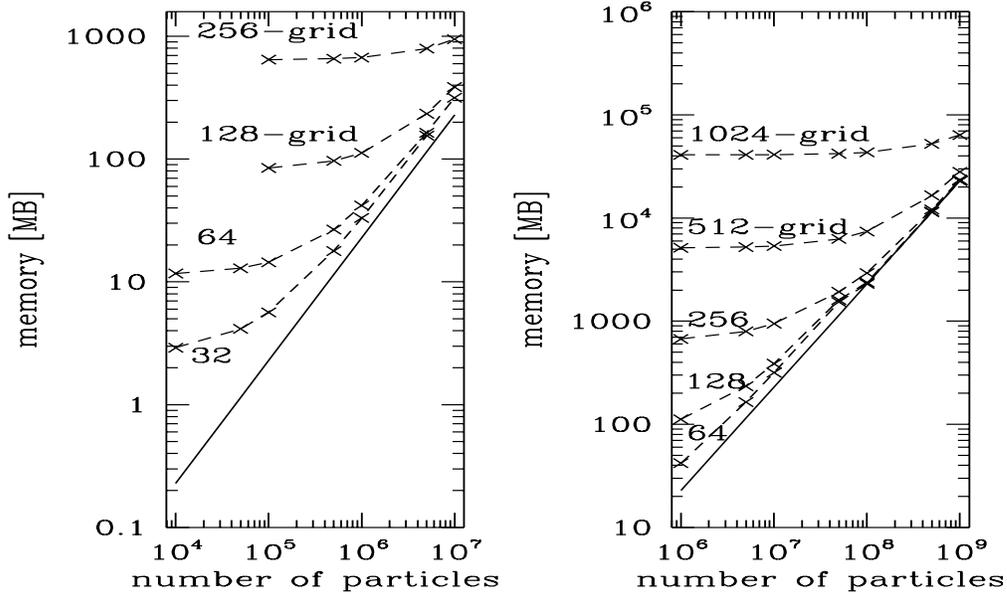}
    \caption{The total memory requirement of {\sc Superbox} for
      different grid-resolutions as function of the
      particle-number $N_{\rm p}$ (straight line is the memory
      requirement of the particle array alone).}
    \label{mem}
  \end{center}
\end{figure}

\noindent
The memory requirement of {\sc Superbox} scales both linearly
with the particle number, $N_{\rm p}$, and with the number of
grid-cells, $N_{\rm gc}=N^3$.  For the particle array, 24~byte
per particle (4~bytes per phase-space variable) are needed.  The
total amount of memory for different particle numbers and
grid-resolutions is shown in Fig.~\ref{mem}.

\noindent
The memory required for a {\sc Superbox} calculation can be
estimated from 
\begin{eqnarray}
  {\rm memory_{grid}} & = & \left(5 \cdot N^{3} + N (2N)^{2} +
  (N+1)^{3} \right) \times 4 \ {\rm byte},  \label{f5.1} \\
  {\rm memory_{particles}} & = & \left( 8 \cdot N_{\rm p} +
  {\rm overhead} \right) \times 4 \ {\rm byte}.  \label{f5.1a}
\end{eqnarray}
The first term on the right-hand side of Eq.~\ref{f5.1}
comes from the 5~grids (Section~\ref{sec:grid}).  The second term
comes from the array needed for the FFT, and the last term comes
from the array required to store the Fourier-transformed Green's
function.  The memory usage of the grids is independent of the
number of galaxies, since they are treated consecutively in the
same grid-arrays.  The 'overhead' contains the arrays of the {\sc
  Superbox} parameters (e.g. grid sizes), the centre of mass and
density, and output data such as energies, angular momenta and
Lagrangian radii, for each galaxy.  Increasing the number of
galaxies, while keeping the total particle and grid number
constant, reduces the relative overhead contribution
(Eq.~\ref{f5.1a}) slightly due to these large arrays scaling
only with the number of particles per galaxy. 

\subsection{CPU-time}
\label{sec:cpu}

\begin{figure}[h!]
  \begin{center}
    \leavevmode
    \epsfxsize=14cm
    \epsfysize=10cm
    \epsffile{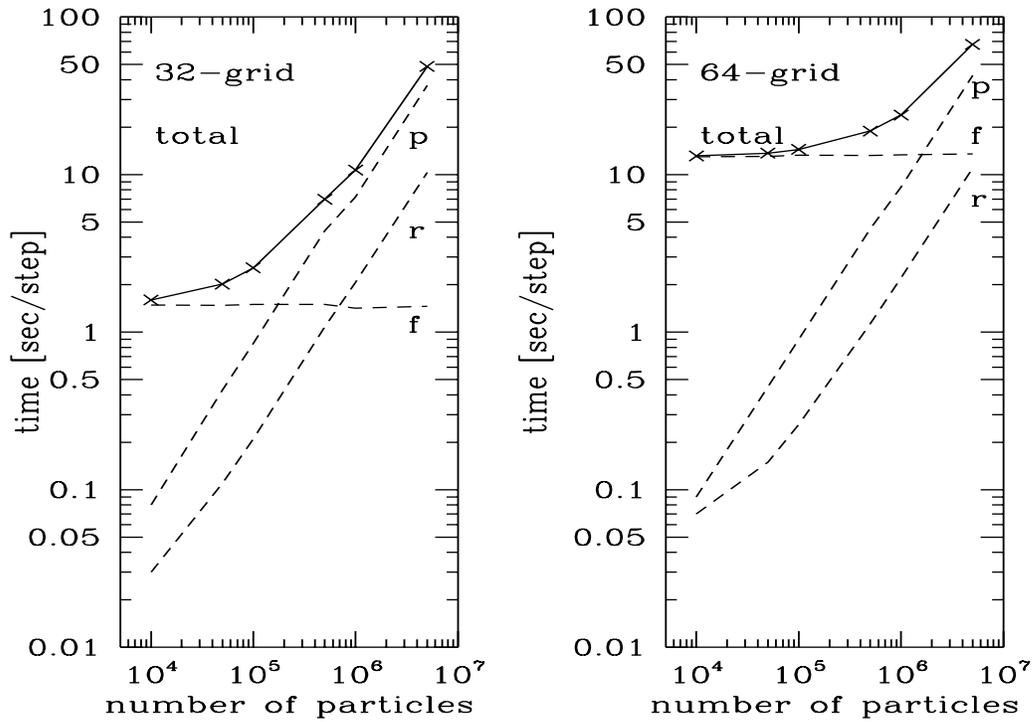}
    \caption{Contributions to the CPU time, $t_{\rm CPU}$, 
      by different routines,
      and total CPU-time per step for different particle numbers
      and grid resolutions (r denotes the {\sc Getrho}-routine, f
      the {\sc fft}-routine and p the {\sc Pusher}-routine).}
    \label{cpu}
  \end{center}
\end{figure}

\noindent
The time-step cycle of {\sc Superbox} is divided into three main
routines (Fig.~\ref{super}).  Firstly, in {\sc Getrho} the mass
density of the grids is calculated.  Secondly, the {\sc FFT} routine
computes the potential on the grids.  Thirdly, the {\sc Pusher}
routine contains the force calculation, the position and velocity
updating, and collection of the output data.  These three
routines need about 99~per cent of the total CPU time.
Fig.~\ref{cpu} shows the amount of CPU-time per step, $t_{\rm
  CPU}$, needed for the different routines.  All data are derived
on a Pentium II 400MHz processor with Linux as the operating
system and the GNU--g77 Fortran Compiler (with options $-$O3
$-$m486) for a model single galaxy. 
\begin{figure}[h!]
  \begin{center}
    \leavevmode
    \epsfxsize=12cm
    \epsfysize=9cm
    \epsffile{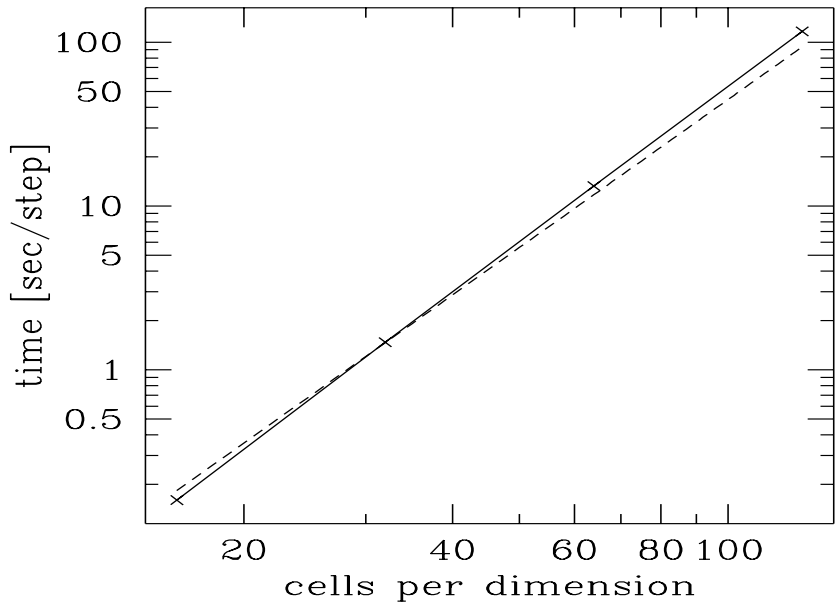}
    \caption{CPU-time per step used by the {\sc fft}-routine
      for different grid-resolutions (solid line). The dashed
      line shows $t \propto N_{\rm gc}=N^{3}$. }
    \label{cpu2}
\end{center}
\end{figure}
For the {\sc Getrho} and {\sc Pusher} routines, $t_{\rm CPU}
\propto N_{\rm p}$, while for the {\sc FFT} routine, $t_{\rm CPU}
\propto N_{\rm gc} \log_{10} N_{\rm gc}$ ($N_{\rm gc} = N^{3}$;
Fig.~\ref{cpu2}), which is by far the dominant contribution.  The
resultant CPU time per step for galaxy $i$ is:
\begin{eqnarray}
  t_{{\rm CPU},i} & = & \alpha\,N_{p,i} + \beta \times 5\,N_{\rm
    gc}{\rm log}_{10}N_{\rm gc}\ .  
\label{eqn:cpu}
\end{eqnarray}
As another example, for a Pentium~200MHz MMX, $t_{{\rm CPU}} \approx
20$~sec/step for $N_{\rm p}=10^6$ particles in one galaxy and
$N_{\rm gc}=64^3$ grids. 

The CPU-time needed for a computation with $N_{\rm gal}$ galaxies
is derived by adding the individual times,
\begin{eqnarray}
  t_{\rm CPU, tot} = \sum_{i=1}^{N_{\rm gal}}\,t_{{\rm CPU},i}\ .
\label{eqn:cpu_tot}
\end{eqnarray}

\section{Comparison with other Codes}
\label{sec:fin_valid}

\noindent
Demonstration of the conservation of energy and angular momentum
is a necessary but not sufficient condition for validating
any particle-mesh code.  Additional validation of the code comes from
an inter-comparison of results obtained with entirely different
numerical techniques.  Such a comparison is available in the
study of the tidal dissolution of small satellite galaxies
orbiting in an extended Galactic dark halo (Klessen \& Kroupa
1998). Such an application of {\sc Superbox} is rather extreme in 
that the jump in resolution from the middle grid (0.29~kpc/cell
length) containing the small satellite to the outermost grid
(25~kpc/cell length) containing the 'local universe' is very
large. 

In that study, two {\sc Superbox} computations are compared with
calculations done using a direct but softened N-body integrator
on the special hardware device named {\sc Grape}~3 connected with
a SUN-Ultra~20.  The {\sc Superbox}
calculation uses $N_{\rm gc}=32^3$ grids, has $3\times10^5$
satellite particles, and treats the live Galactic dark halo with
$10^6$ particles.  The {\sc Grape} computation uses $1.3 \times
10^5$ satellite particles with the Galactic dark halo being an
analytic isothermal potential.  Both calculations proceed for
many thousands of time steps, corresponding to a time interval of
many~Gyr, until beyond disruption of the satellite.  The results
are in nice agreement, yielding essentially the same behaviour of
the satellite, apart from expected (small) differences in exact
disruption time.  Of interest is also a comparison in CPU times
required for the calculations, keeping in mind the very different
number of particles used: with {\sc Superbox} it takes~0.57 days
on an IBM~RISC/6000 workstation to compute 1~Gyr using $1.3
\times 10^6$ particles, and it takes 0.36~days with {\sc Grape}
using $ 1.3 \times 10^5 $ particles (with an ${\mathcal O}(N_{\rm
  p}^2)$ scaling).  

In the above calculations, the satellite mass is too small for
the orbit to be affected by dynamical friction.  A comparison of
the orbital decay through dynamical friction of a massive
satellite is of interest with analytical estimates, because this
allows an altogether independent verification if {\sc Superbox}
treats collective behaviour on larger scales correctly.  This is
discussed in Section~\ref{sec:dynfric}.


\section{Dynamical Friction}
\label{sec:dynfric}

\noindent
An object moving through a homogeneous distribution of much
lighter masses induces a trailing over-density. This over-density
attracts the object which, as a result, is decelerated
(Chandrasekhar 1943). 

The orbital decay of a satellite galaxy or globular cluster
orbiting within a larger galaxy is an astrophysical process
relevant, for example, to the rate at which a satellite
population is depleted and to the built-up of galactic bulges.
In the Local Group, dynamical friction affects the orbit of the 
Magellanic Clouds, which is important for tracing their origin
(see Westerlund 1997 and Kroupa \& Bastian 1997 for
summaries). It may also change the orbit of the Sagittarius
dwarf spheroidal galaxy (Ibata \& Lewis 1998; Gomez-Flechoso et
al.\ 1999) and other nearby 
companions, depending on their masses. Understanding and proper
application of dynamical friction is thus a very important
issue. 

The question if Chandrasekhar's formula (Eq.~\ref{eqn:eom1} 
below) can be applied to orbits within finite and inhomogeneous
density distributions has been the issue of a significant debate
throughout the 1970's and the 1980's. The result is that for
satellite masses that are relatively small compared to the large
galaxy, the formula is a good approximation (Bontekoe \& van
Albada 1987; Zaritsky \& White 1988; Velazquez \& White 1999).
Massive satellites induce non-local perturbations in the larger
galaxy that are not taken account of in the derivation of
Eq.~\ref{eqn:eom1}, and in the collision of two galaxies of
comparable mass analytical estimates become intractable, and the
self-consistent numerical experiment must be resorted to
(e.g. Madejski \& Bien 1993). In order to do this we must,
however, have reason to trust the numerical results. 

To demonstrate the reliability of our results, we compare the 
orbital evolution of a satellite galaxy orbiting in an extended
dark halo computed with the fully self-consistent {\sc
  Superbox}-code, with the evolution resulting from the by now
well-established Chandrasekhar approximation (Eq.~\ref{eqn:eom1}).

\subsection{Calculations with {\sc Superbox}}
\label{sec:superb_dynfr}

\noindent
For the numerical experiment a compact spherical satellite galaxy
is injected into an extended isothermal dark parent halo.  The
numerical rendition of both systems is described in detail in
Kroupa (1997), and only a short description is given here. 

The parent halo is assumed to extend to $R_{\rm c} = 250$~kpc
with a core radius $\gamma=5$~kpc and a total mass $M_{\rm
  halo} = 2.85 \times 10^{12} \, M_\odot$, corresponding to a
circular velocity $V_{\rm c} = 220$~km/s.  A particle takes
588~Myr to cross the 33~per cent mass diameter.  The inner,
middle and outer grids are centred on the density maximum and
have edge lengths of~50, 188~and 700~kpc, respectively.  The
model is allowed to relax to virial equilibrium, after which
state it has a slightly more compact configuration with a
circular velocity at a radius of 150~kpc of 245km/s,
corresponding to a mass within that radius of about $2.1 \times
10^{12}\, M_\odot$.  Due to a mild radial orbit instability the
halo is also slightly prolate. 

Plummer density distributions are taken for the satellites, each
with a Plummer radius $R_{\rm pl} = 3$~kpc and a cutoff radius of
15~kpc.  The innermost and middle grids are centred on the
satellite's density maximum and have edge lengths of~10
and 40~kpc, respectively.  The outer grid is the same as for the
halo.  Three satellite masses are used, $M_{\rm sat} = 1 \times
10^{10}, 7 \times 10^{10}\, {\rm and}\, 5 \times 10^{11}\,
M_\odot$.  The respective crossing times of the 33~per cent mass
diameter are~$t_{\rm cr,33} = 84$, 32 and 12~Myr. 

For the calculation, $N = 32$ grid cells per dimension are used
(only 28~being active, see Section~2.3), with $10^6$ particles in
the parent halo, and $3 \times 10^5$ satellite particles.  For the
halo the resolution is thus 1.79~kpc/cell-length for $r \leq
25$~kpc, 6.71~kpc/cell-length for 25~kpc $< r \leq 94$~kpc 
and 25~kpc/cell-length for 94~kpc $ < r \leq 350$~kpc.  For the
satellite the resolution is 0.36~kpc/cell-length for $r_{\rm
  s} \leq 5$~kpc, 1.43~kpc/cell-length for 5~kpc $ < r_{\rm s}
\leq 20$~kpc and 25~kpc/cell-length for 20~kpc $ < r_{\rm s} \leq
350$~kpc, where $r_{\rm s}$ is the distance from the satellite's
density maximum.  The integration step size is $\Delta t = t_{\rm
  cr,33} / 75$.

The satellite is allowed to relax to virial equilibrium before
being placed into the parent halo at an initial radius $r(0) =
|{\bf r}(t=0)| = 150$~kpc with an initial velocity $v(0) = |{\bf
  v}(t=0)| = 245$~km/s perpendicular to the radius vector, ${\bf
  r}$. 

The satellite's galactocentric distance is $r(t) = |{\bf r}_{\rm
sat}(t) - {\bf r}_{\rm halo}(t)|$, and its velocity is $v(t) =
|{\bf v}_{\rm sat}(t) - {\bf v}_{\rm halo}(t)|$, where ${\bf
  r}_{\rm sat}(t)$ and ${\bf r}_{\rm halo}(t)$ are the position
vectors of the density maxima of the satellite and halo,
respectively, and ${\bf v}_{\rm sat}(t)$ and ${\bf v}_{\rm
  halo}(t)$ are the velocity vectors of the satellite's and
halo's density maxima, respectively. 

\subsection{Chandrasekhar friction}
\label{sec:chandra_dynfr}

\noindent
The equation of motion of a satellite in a stationary and rigid 
isothermal parent halo is 
\begin{eqnarray}
  {{\rm d}^2{\bf r}\over {\rm d}t^2} &=& a_{\rm g}\,{\bf r} +
   \eta\, {\bf v},\label{eqn:eom1}\\
   a_{\rm g}&=&-{V_{\rm c}^2 \over \gamma^2+r^2},
   \label{eqn:eom2}\\ 
   \eta &=& -4\pi\,G^2 {\rm ln}\Lambda {\rho(r)M_{\rm sat}\over
     v^3} \left[{\rm erf}(x)-{2\,x\,{\rm e}^{-x^2}\over
       \sqrt{\pi}}\right],\label{eqn:eom3}
\end{eqnarray}
where $a_{\rm g}$ is the acceleration in an isothermal halo,
and $\eta$ is the deceleration due to dynamical friction, in
which $G = 4.499 \times 10^{-3}\, {\rm pc}^3/ (M_\odot\, {\rm
  Myr}^2)$~is the gravitational constant, $\rho(r) = V_{\rm
  c}^2 / \left[ 4 \pi {\rm G} \left( \gamma^2 + r^2 \right)
\right]$ is the halo density and erf$(x)$ is the error function
with $x = v / V_{\rm c}$.  A derivation of Chandrasekhar's
dynamical friction formula can be found in Binney \& Tremaine
(1987). 

The numerical value of the Coulomb logarithm, ln$\Lambda = b_{\rm 
max} / b_{\rm min}$, is somewhat ill-defined.  It is calculated by
integrating particle deflections over impact parameters ranging
from a minimum ($b_{\rm min}$) to a maximum ($b_{\rm max}$)
effective value.  The minimum impact parameter is taken here to
be $b_{\rm min} = 2.7$~kpc.  The maximum impact parameter, however,
is something like the distance over which $\rho(r)$ falls off
significantly, and is thus less-well defined.  An upper limit is
$b_{\rm max} \approx r(t=0)=150$~kpc.  It can also be estimated by
assuming $\rho(r+b_{\rm max}^+) = 0.5\,\rho(r)$.  With $r =
150$~kpc, $b_{\rm max}^+ = 62$~kpc.  If, on the other hand,
$\rho(r-b_{\rm max}^-) = 2 \, \rho(r)$, then with $r = 150$~kpc,
$b_{\rm max}^- = 44$~kpc.  Once the satellite orbit decays to $r
= 30$~kpc, then $b_{\rm max}^+ = 13$~kpc, while $b_{\rm max}^- =
9$~kpc.  The Coulomb logarithm thus lies in the range ln$\Lambda
\approx$~1 to~4, and the above argument suggests that it may be a
monotonically decreasing function of declining $r(t)$. 

Apart from the somewhat arbitrary value of the Coulomb logarithm,
the analytic derivation of Eq.~\ref{eqn:eom3} assumes a
Maxwellian velocity distribution in the halo, neglects the self
gravity of the wake induced by the satellite's gravitational
focusing, and the motion of the deflected halo particles is
assumed to be governed only by the satellite -- the gravitational
field of the parent halo is neglected.  Nevertheless,
Eq.~\ref{eqn:eom3} has been shown to provide reliable
results provided $M_{\rm sat} / M_{\rm halo} \simless 0.2$ and
$\gamma < r(t) < R_{\rm c}$ (Binney \& Tremaine 1987 and
references therein).  The {\sc Superbox} model described in
Section~\ref{sec:superb_dynfr} conforms with these
restrictions. 

For the comparison with {\sc Superbox}, $V_{\rm c} = 245$~km/s,
$\gamma = 3$~kpc, and $M_{\rm sat} = 1 \times 10^{10}, \ 7 \times
10^{10}$ and $5\times10^{11}\,M_\odot$.  Eq.~\ref{eqn:eom1} is
rewritten to four coupled first-order differential equations, and
integrated numerically using the Runga-Kutta method, with a
sufficiently small step size to ensure stability of the solution.  

\subsection{Results}
\label{sec:res_dynfr}

\begin{figure}[h!]
  \begin{center}
    \leavevmode
    \epsfxsize=12.0cm
    \epsfysize=16.5cm
    \epsffile{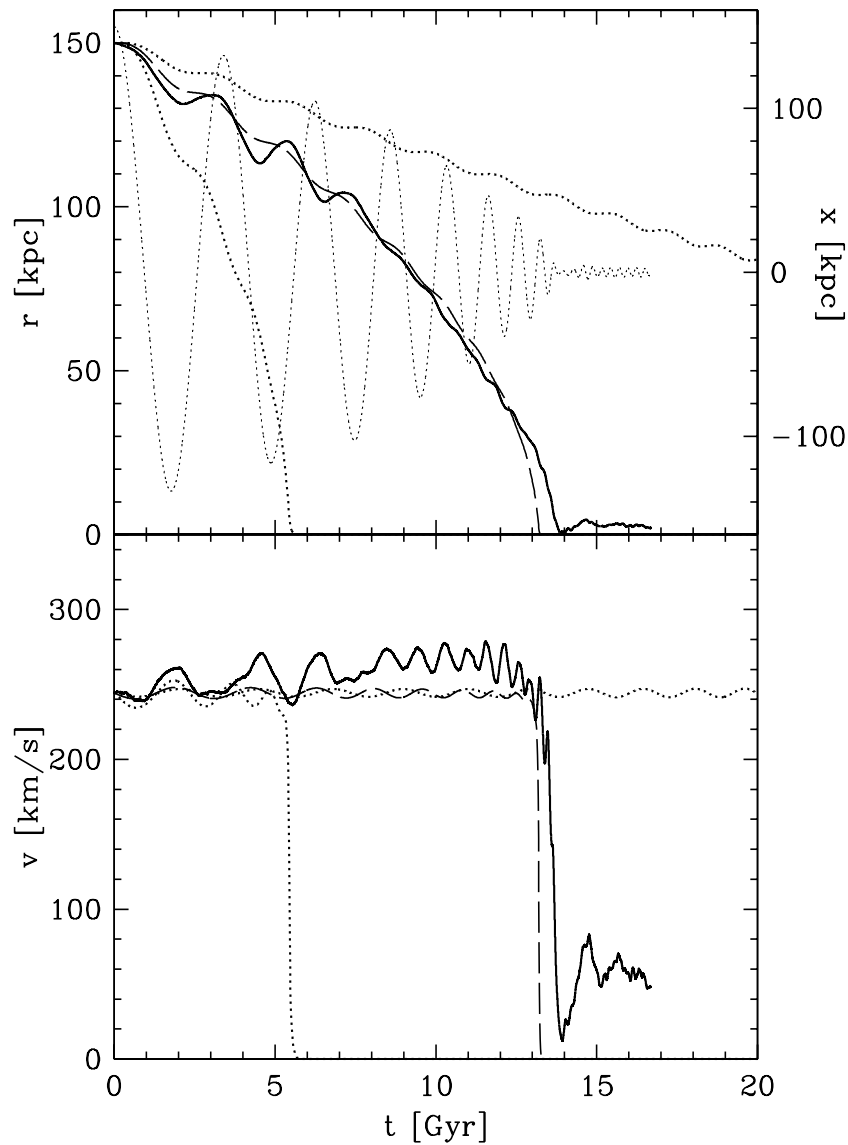}
    \caption{
      The time evolution of the satellite's galactocentric
      distance, $r$ (upper panel), and velocity, $v$ (lower
      panel).  The satellite has a mass $M_{\rm sat} = 7 \times
      10^{10}\, M_\odot$ and orbits in an extended isothermal
      dark halo.  The {\sc Superbox} result is shown as the thick
      solid curve.  The other (theoretical) curves result from 
      numerically integrating Eq.~\ref{eqn:eom1} assuming
      ln$\Lambda = 4$ (lower dotted curve in top panel),
      ln$\Lambda = 1.6$ (long-dashed curve) and ln$\Lambda(t) =
      {\rm ln}(0.049\, r(t) / b_{\rm min})$ with $b_{\rm min} =
      2.7$~kpc (upper dotted curve in top panel).  The {\sc 
      Superbox} evolution of $x(t) = x_{\rm sat}(t) - x_{\rm
      halo}(t)$ is plotted as the thin dotted line in the upper 
      panel.} 
    \label{fig:vr1}
  \end{center}
\end{figure}

\noindent
Since the Coulomb logarithm is ill defined, it is necessary to
first calibrate the numerical solution to Eq.~\ref{eqn:eom1}
using one {\sc Superbox} calculation. Fig.~\ref{fig:vr1} shows
the decay of the orbit of the satellite with $M_{\rm sat} = 7
\times 10^{10}\,M_\odot$ according to {\sc Superbox} and
Eq.~\ref{eqn:eom1} under different assumptions for
ln$\Lambda$. 

For the solution shown as the lower-dotted-$r(t)$ curve in
Fig.~\ref{fig:vr1}, ln$\Lambda = 4$. The {\sc Superbox} result
requires a smaller Coulomb logarithm.  Tests show that
ln$\Lambda = 1.6$ leads to good agreement.  To show a possible
upper limit to the solution of the equation of motion, the case
ln$\Lambda = (0.049\, r(t)/b_{\rm min})$ is plotted as the
upper-dotted-$r(t)$ curve.  For this case, ln$\Lambda = 1$
initially, and it decreases with $r(t)$.  The figure also shows
the time-varying distance along the x-axis between the density
maxima of the satellite and the halo.  The damped oscillation is
nicely evident, and the period of the orbit is initially 3.2~Gyr.
The lower panel of Fig.~\ref{fig:vr1} plots the solutions for
$v(t)$.  The satellite in the {\sc Superbox} calculation retains
an approximately constant velocity, as is expected from the
theory of dynamical friction, the solutions from which are
plotted with different lines corresponding to the cases discussed
above. 

\begin{figure}[h!]
  \begin{center}
    \leavevmode
    \epsfxsize=12.0cm
    \epsfysize=16.3cm
    \epsffile{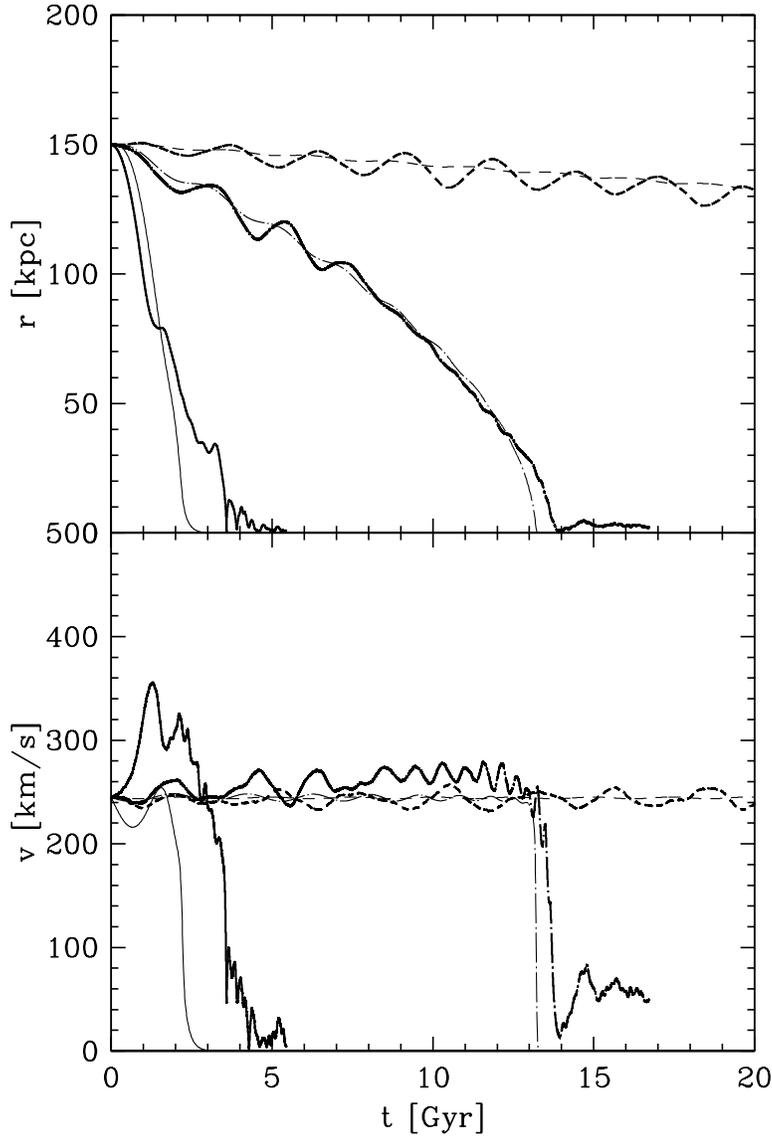}
    \caption{ The time evolution of the satellite's
      galactocentric distance, $r$ (upper panel), and velocity,
        $v$ (lower panel). The {\sc Superbox} satellite has a
        mass $M_{\rm sat} = 5 \times 10^{11} \ ,M_\odot$ (solid
        curve), $7 \times 10^{10} \, M_\odot$ (dash-dotted curve)
        and $1 \times 10^{10} \, M_\odot$ (dashed curve), and
        orbits in an extended isothermal dark halo. The case
        discussed in Fig.~\ref{fig:vr1} is identical to the
        present {\sc Superbox} case plotted as the dash-dotted
        curve.  Eq.~\ref{eqn:eom1}, with ln$\Lambda = 1.6$,
        produces the corresponding solutions shown as thin
        lines.}  
    \label{fig:vr2}
  \end{center}
\end{figure}

The overall conclusion is that the {\sc Superbox} result is
consistent with theoretical expectations, provided
ln$\Lambda = 1.6$.  That this finding extends to other satellite
masses is shown in Fig.~\ref{fig:vr2}. In all cases, the
satellites have lost less than 20~per cent of their mass by the
end of the calculation. 

It is notable that one single value for the Coulomb logarithm
applies to the whole radial range as well as to all three
satellites, suggesting that non-local perturbations of the
extended halo do not significantly alter the orbit of the
satellites.  This finding is in nice agreement with the results
for ln$\Lambda$ by Bontekoe \& van Albada (1987) and Velazquez \&
White (1999), and support their finding that Chandrasekhar's
approximation yields a useful description of orbital evolution if
ln$\Lambda \approx 1.5$ is used. 

Further calculations with more grid cells and a more detailed
study will be needed to show if the oscillations in $r(t)$ and
$v(t)$ seen in Figs.~\ref{fig:vr1} and~\ref{fig:vr2}, as well as
in similar figures in the literature (e.g. fig.~3 in Bontekoe \&
van Albada 1987) are physical, i.e. if they are manifestations of
the dependency of $\eta$ (Eq.~\ref{eqn:eom3}) on $r$ and
$v$.


\section{Conclusions}
\label{sec:conc}

\noindent
{\sc Superbox} is an unconventional particle-mesh code that uses
moving sub-grids to track and resolve high-density peaks in
the particle distribution.  This extension avoids the limitation
in spatial resolution encountered with standard particle mesh
codes that employ only a single grid.  The code is efficient in
that the computational overhead is kept as slim as possible.  The
code is also memory efficient by using only one set of grids to
treat galaxies in succession. 

In this paper we have explained the algorithms used for force
evaluation, grid positioning and orbit integration.  The
nearest-grid-point scheme (NGP) is used to compute the density
on a grid.  The potential is obtained by FFT, from which
accelerations at the position of each particle within a cell are
calculated using second-order central-difference quotients with
linear interpolation to the position of particles inside that
cell. Such a scheme has been identified as a special higher-order
NGP scheme and is competitive with the cloud-in-cell (CIC) scheme
regarding the precision and continuity of the forces across grid 
boundaries.  Together with the nested sub-grids, this leads to
precise and reliable force calculations at the location of each
particle.  The sub-grids are positioned at each time-step on the
density maximum or at the centre of mass of a galaxy, the former
method being the more useful.  For orbit integration the
leap-frog scheme is used.  Conservation of energy and angular
momentum is excellent, and numerical relaxation is significantly
longer than a Hubble time for a galaxy with a crossing time of
about $10^8$~yr and particle numbers exceeding $10^5$.
CPU-timing and memory consumption are such that calculations of
galaxy encounters involving many $\times 10^6$ particles are
possible on present-day desk-top computers with 20MB RAM or more
(see Fig.~\ref{mem}).  On the other hand, it is clear that using
the resources of massively parallel supercomputers will
significantly increase the capabilities of {\sc Superbox}. 

The most serious limitation of {\sc Superbox} lies in the
inability to resolve newly formed self-gravitating systems,
because the grid structure is decided on at the beginning of a
computation.  {\sc Superbox} has presently no scheme to allow the
inclusion of additional grids during a calculation.  Such objects
can form in tidal arms and may be the precursors of some dwarf
satellite galaxies.  This limitation will be reduced as the
number of grids that can be used increases owing to advances in
computer technology. 

Current and future developments of {\sc Superbox} would include
for example incorporating an individual time-step scheme per grid
and galaxy.  We intend to parallelise the code to run on CRAY~T3E
super-computers.  Furthermore, additional grid-levels will be
introduced to reduce the jump in resolution between the present
middle and outer grids, and/or to further increase the resolution
of the central regions of a galaxy.  Finally, the current
sticky-particle algorithm of {\sc Superbox} can be extented
(e.g.\ to include star formation). 

Problems that are now being tackled with {\sc Superbox} include
computations of the tidal interactions of satellite and disk galaxies
(J.M. Penarrubia, in preparation), and the possible formation of
spheroidal satellite galaxies from stellar super clusters (Fellhauer
et al.). Fellhauer \& Kroupa (2000) report on a study of the dynamical
evolution of clusters of dozens to hundreds of young massive and
compact star clusters (i.e. {\it stellar super clusters}) in a tidal
field, in an attempt to understand the evolution of such objects
observed in HST images of the Antennae galaxies (see e.g. Lan\c{c}on
et al. 2000).  Also, the parameter survey of dwarf galaxies orbiting
in a massive dark halo is being completed (Kroupa, in preparation), to
identify regions in parameter space that may lead to dSph-like systems
without dark matter.  Furthermore, collapse calculations are being
performed to study the properties of the resulting object after
violent relaxation and secular relaxation thereafter (Boily, in
preparation). 


\noindent{\bf Acknowledgements}
\noindent

P.K. and H.B. acknowledge financial support by SFB 328 and C.B,
by SFB439 (at the University of Heidelberg) from the German
Science Foundation (DFG).  Support and advice on computational
aspects of HLRS Stuttgart is gratefully acknowledged. \\
{\sc Superbox} is available at\\
\vspace*{-1cm}
\begin{verbatim}
ftp:\\ftp.ari.uni-heidelberg.de/pub/mike/super.tar.gz.
\end{verbatim}



\begin{thebibliography}{99} 

\bibitem{Ba86} Barnes J.E., Hut P., 1986, Nature, {\bf 324}, 446

\bibitem{Bi91} Bien R., Fuchs B., Wielen R., 1991, in The CP90
               Europhysics Conference on Computational Physics, 
               Serie A, {\bf 228}, 3

\bibitem{Bi87} Binney J., Tremaine S., 1987, Galactic Dynamics,
               Princeton University Press, New Jersey

\bibitem{Bo87} Bontekoe Tj.R., van Albada T.S., 1987, MNRAS, {\bf
               224}, 349 

\bibitem{Ch43} Chandrasekhar S., 1943, ApJ, {\bf 97}, 255

\bibitem{Co99} Couchman H.M.P., 1999, to appear in Riffert H.,
               Werner K. (eds.), Computational Astrophysics, The
               Journal of Computational and Applied Mathematics
               (JCAM), Elsevier Press, Amsterdam

\bibitem{Da97} Dav\'e R., Dubinski J., Hernquist L., 1997, NewA,
               {\bf 2}, 277

\bibitem{Ea78} Eastwood J.W., Brownrigg D.R.K., 1978, J.\ Comput.\
               Phys., {\bf 32}, 24

\bibitem{Fe96} Fellhauer M., 1996, diploma thesis, Univ.\ of
               Heidelberg 

\bibitem{Fe2k} Fellhauer M., Kroupa P., 2000, in ASP Conf. Series
               Vol. 211, Massive Stellar Clusters,
               ed. A. Lan\c{c}on \& C.M. Boily (San Francisco:
               PASP), 241  
 
\bibitem{Go99} Gomez-Flechoso M.A., Fux R., Martinet L., 1999,
               A\&A, {\bf 347}, 77

\bibitem{Ho81} Hockney R.W., Eastwood J.W., 1981, Computer
               Simulations Using Particles, McGraw-Hill 

\bibitem{Ho70} Hohl F., 1970, NASA Technical Report {\bf R-343} 

\bibitem{Ib98} Ibata R.A., Lewis G.F., 1998, ApJ, {\bf 500}, 575 

\bibitem{Kl98} Klessen R., Kroupa P., 1998, ApJ, {\bf 498}, 143 

\bibitem{Kra97} Kravtsov A.V., Klypin A.A., Khokhlov A.M., 
                1997, ApJS, {\bf 111}, 73

\bibitem{Kr97} Kroupa P., 1997, NewA, {\bf 2}, 139

\bibitem{KrB97} Kroupa P., Bastian U., 1997, NewA, {\bf 2}, 77

\bibitem{La2k} Lan\c{c}on A., Boily C.M. (eds.), 2000, ASP
               Conf. Series Vol. 211, Massive Star Clusters, San
               Francisco: PASP, pp. 330 
 
\bibitem{Ma93} Madejski R., Bien R., 1993, A\&A, {\bf 280}, 383

\bibitem{Pe97} Pearce F.R., Couchman H.M.P., 1997, NewA, {\bf 2},
               411 

\bibitem{Pr86} Press W.H., Teukolsky S.A., Vetterling W.T.,
               Flannery B.P., 1986, Numerical Recipes in Fortran,
               Cambridge University Press, Cambridge

\bibitem{Se87} Sellwood J.A., 1987, ARA\&A, {\bf 25}, 151

\bibitem{Sp71} Spitzer L., Hart M.H., 1971, ApJ, {\bf 164}, 399

\bibitem{Sp99} Spurzem R., 1999, Journal of Computational and
               Applied Mathematics, {\bf 110}, Elsevier, in press 

\bibitem{St69} Standish E.M., Aksnes K., 1969, ApJ, {\bf 158},
               519 

\bibitem{Ve99} Velazquez H., White S.D.M., 1999, MNRAS, {\bf
               304}, 254 

\bibitem{Vi89} Villumsen J.V., 1989, ApJS, {\bf 71}, 407

\bibitem{Wa92} Wassmer N., 1992, diploma thesis, Univ.\ of
               Heidelberg 

\bibitem{Wa93} Wassmer N., Bien R., Fuchs B., Wielen R., 1993, 
               Astron. Ges. Abstr. Ser., {\bf 9}, 78

\bibitem{We97} Westerlund B.E., 1997, The Magellanic Clouds,
               Cambridge University Press, Cambridge

\bibitem{We79} Werner H., Schabach R., 1979, Praktische
               Mathematik II, Springer Verlag

\bibitem{Za88} Zaritsky D., White S.D.M., 1988, MNRAS, {\bf
               235}, 289 

\end{thebibliography}
\end{document}